\documentclass[%
 reprint,
superscriptaddress,
 amsmath,amssymb,
 aps,
 prl
]{revtex4-1}

\usepackage{graphicx}
\usepackage{dcolumn}
\usepackage{bm}
\usepackage{color}

\usepackage{pdfpages}
\usepackage{pgffor}

\makeatletter
\AtBeginDocument{\let\LS@rot\@undefined}
\makeatother

\begin{document}

\preprint{APS/123-QED}

\title{Towards the complete phase profiling of attosecond wave packets}

\author{Jaco Fuchs}
\email{jafuchs@phys.ethz.ch}
\affiliation{Department of Physics, Eidgenössische Technische Hochschule Z\"urich, Z\"urich, Switzerland}

\author{Nicolas Douguet}%
\affiliation{Department of Physics, Kennesaw State University, Marietta, Georgia, USA}
\affiliation{Department of Physics, University of Central Florida, Orlando, Florida, USA}

\author{Stefan Donsa}
\affiliation{Institute of Theoretical Physics, Vienna University of Technology, Vienna, Austria, EU}

\author{Fernando Mart\'{\i}n}
\affiliation{Departamento de Qu\'imica Modulo 13, Universidad Aut\'onoma de Madrid, 28049 Madrid, Spain, EU}
\affiliation{Condensed Matter Physics Center (IFIMAC), Universidad Autonoma de Madrid, 28049 Madrid, Spain, EU}
\affiliation{Instituto Madrile\~no de Estudios Avanzados en Nanociencia (IMDEA-Nano), 28049 Madrid, Spain, EU}

\author{Joachim Burgd\"orfer}
\affiliation{Institute of Theoretical Physics, Vienna University of Technology, Vienna, Austria, EU}

\author{Luca Argenti}
\affiliation{Department of Physics, University of Central Florida, Orlando, Florida, USA}
\affiliation{CREOL, University of Central Florida, Orlando, Florida 32186, USA}

\author{Laura Cattaneo}
\affiliation{Department of Physics, Eidgenössische Technische Hochschule Z\"urich, Z\"urich, Switzerland}

\author{Ursula Keller}
\affiliation{Department of Physics, Eidgenössische Technische Hochschule Z\"urich, Z\"urich, Switzerland}

\date{\today}

\begin{abstract}
Realistic attosecond wave packets have complex profiles that, in dispersive conditions, rapidly broaden or split into multiple components. Such behaviors are encoded in sharp features of the wave packet spectral phase. Here, we exploit the quantum beating between one- and two-photon transitions in an attosecond photoionization experiment to measure the photoelectron spectral phase continuously across a broad energy range. Supported by numerical simulations, we demonstrate that this experimental technique is able to reconstruct sharp fine-scale features of the spectral phase, continuously as a function of energy and across the full spectral range of the pulse train, thus beyond the capabilities of existing attosecond spectroscopies. In a proof-of-principle experiment, we retrieve the periodic modulations of the spectral phase of an attosecond pulse train due to the individual chirp of each harmonic.
\end{abstract}

\maketitle

Attosecond photoionization time delays provide a precise timing of electronic motion in atoms~\cite{Schultze2010, Isinger2017, Cirelli2018a}, molecules~\cite{Sansone2010, Vos2018} and solids~\cite{Cavalieri2007,Locher2015,Kasmi2017,Ossiander2018}. Defined as group delay difference between two electron wave packets, they set benchmarks for the most advanced quantum simulations~\cite{Pazourek2012, Marante2017, Jimenez-Galan2016}. However, as group delays are given by the first-order expansion of the spectral phase $\varphi (E)$, they cannot characterize the full wave packet evolution. Indeed, dynamical aspects more complex than a simple delay, such as changes in the wavepacket envelope shape, can only be reconstructed if the energy-dependent spectral phase is measured in full. In particular, strong and sharp variations of $\varphi (E)$ are key to the most intricate wave packet dynamics~\cite{Gruson2016, Kotur2016, Cirelli2018a, Stoos2018, Polli2010}.

Most experimental techniques currently used to characterize photoionization phases can only retrieve the average value of the group delay across a broad energy range, e.g., the whole attosecond pulse bandwidth in streaking measurements \cite{Itatani2002, Schultze2010, Ossiander2016}, or at discrete energies spaced by twice the probe frequency, in the RABBITT (reconstruction of attosecond beatings by interference of two-photon transitions) scheme \cite{Paul2001, Muller2002}. In these techniques, therefore, rapid phase variations with energy are typically lost. A few interferometric schemes have been proposed to resolve sharp spectral features: by dispersing broad RABBITT sidebands~\cite{Gruson2016, Busto2018, Cirelli2018a}, by scanning the probe frequency across the feature~\cite{Kotur2016, Barreau2019}, or by employing bi-circular attosecond pulse trains~\cite{Donsa2019a}. 
Even these more advanced schemes, however, are sensitive only to the difference of the spectral phase between two isolated harmonics, and hence they can characterize the wave packet profile in more detail only under the ad-hoc assumption that the harmonics are Fourier limited. The question arises, therefore, whether sharp phase variations associated with either the impinging light or the electronic structure of the target can be directly observed.

In this work, we demonstrate that the quantum beat between one- and two-photon transitions, formerly referred to as 1-2 quantum beat \cite{Laurent2012, Laurent2013, Loriot2017, Loriot2020}, together with angle-resolved electron spectroscopy, provides direct access to complex structures in the spectral phase of the photoionized electron wave packets, which, to the best of our knowledge, are inaccessible to any other attosecond spectroscopy method. In contrast to the previous methods \cite{Laurent2012, Laurent2013, Loriot2017, Loriot2020}, we enable the 1-2 quantum beat by performing a RABBITT-inspired experiment using an extreme ultraviolet (XUV) attosecond pulse train (APT) with only odd, but spectrally broad, high harmonics. The combination of the 1-2 quantum beat with spectrally broad high harmonics allows us to retrieve phase differences continuously as a function of energy and across the entire bandwidth of the XUV spectrum, i.e., it allows for a complete phase profiling. In a proof-of-principle experiment, we observe periodic oscillations in the phase of electron wave packets generated by photoionization from helium.
Supported by numerical solutions of the time-dependent Schr\"odinger equation (TDSE), we can attribute these phase oscillations to the harmonic chirp of the XUV pulse train inherent to the underlying high harmonic generation (HHG) process. Whereas the harmonic chirp has been successfully quantified for single harmonics~\cite{Shin1999, Shin2001, Mairesse2005, Haessler2012, Ardana-Lamas2016, Kim2013}, the direct observation of the underlying phase modulations across the full spectrum, originally predicted more than 15 years ago ~\cite{Varju2005, Mauritsson2004}, has not been possible so far.

The spectral phase of a photoelectron wave packet created by absorption of one XUV photon comprises two contributions, the Eisenbud-Wigner-Smith (EWS) scattering phase due to half-scattering at the ionic potential \cite{Wigner1955, Smith1960} and the spectral phase of the ionizing light pulse. The spectral phase of photo-emitted electrons, therefore, can either be used to study the EWS scattering phase \cite{Nagele2011a,Klunder2011} or to characterize XUV light pulses \cite{Mairesse2003,Veniard1996}.
\begin{figure}[t]
\includegraphics[width=0.48\textwidth]{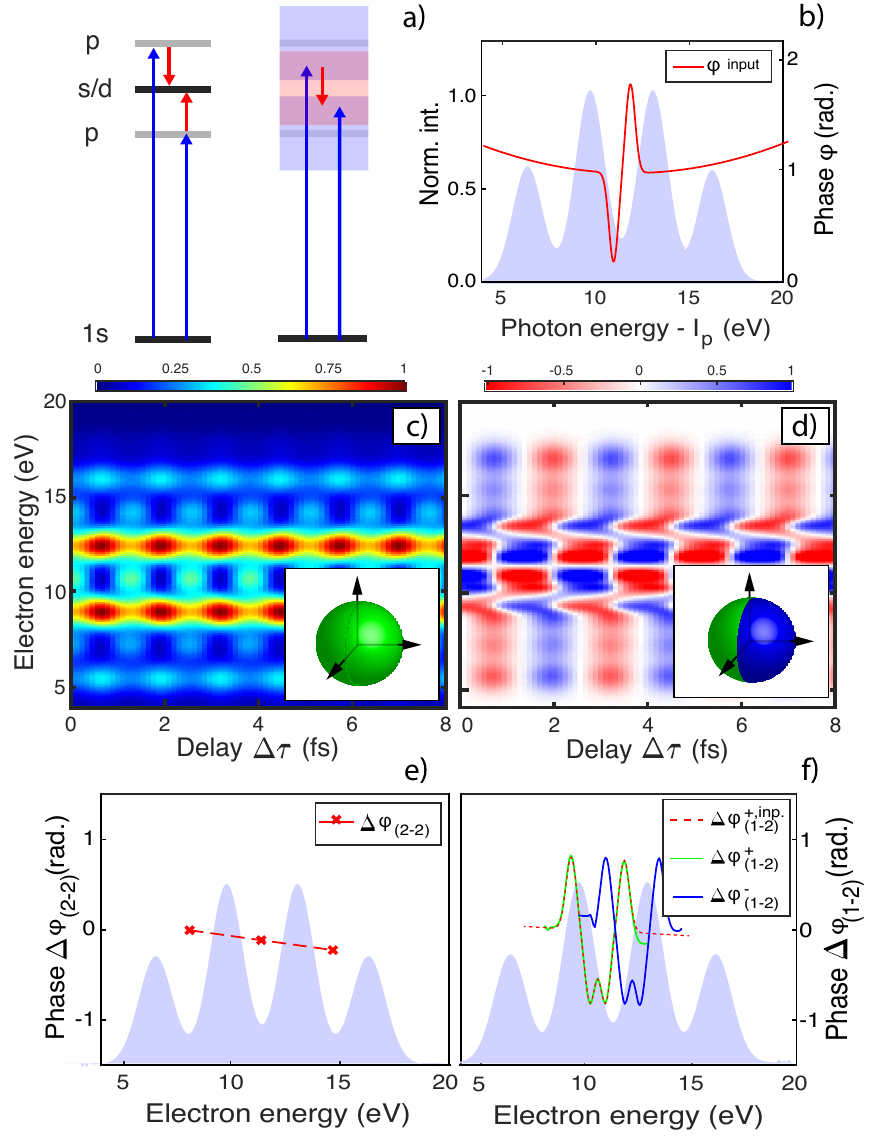}
\caption{\label{fig1} (a) RABBITT (2-2 quantum beat) and the 1-2-quantum beat protocol, schematically. Blue arrows indicate photoionization induced by the XUV and red arrows indicate cc-transitions induced by the IR. (b) Amplitude and phase of the input XUV pulse. (c) RABBITT trace (total yield). The inset indicates the integration over all emission angles. (d) Asymmetry trace extracted from the 1-2 quantum beat. The inset indicates the asymmetry of the angular distribution (difference left-right). (e) Phase difference $\Delta\varphi_{2-2}(E)$ extracted from the RABBITT sidebands. (f) Phase differences $\Delta\varphi_{1-2}^+(E)$ and $\Delta\varphi_{1-2}^-(E)$ extracted from the 1-2 quantum beat method and comparison with $\Delta\varphi_{1-2}^{+,inp.}(E)$ from the input phase.}
\end{figure}
Fig. \ref{fig1}a illustrates the comparison between the RABBITT and the 1-2-quantum beat method described in this letter. Upon XUV photoionization (pump) an IR pulse (probe) promotes continuum-continuum (cc) transitions \cite{Dahlstrom2013,Fuchs2020}. As the pump-probe delay is varied, the photoelectron signal beats as a result of the interference between quantum pathways with the same final energy. The phase of this beating is directly linked to the spectral phase difference between the two interfering quantum paths.
Whereas RABBITT is based on the interference between two different two-photon pathways, i.e., a 2-2-quantum beat, the 1-2 quantum beat method exploits the interference between one-photon and two-photon pathways.

To illustrate the different sensitivity of the two approaches to sharp features in the spectral phase, we first consider an idealized ionization experiment for which we assume that in the energy regime of interest, the atomic ionization cross-section is constant and the EWS and cc-phases are negligible. Under these assumptions, the phase of the ionized electron wave packet directly reflects the phase of the XUV spectrum. The XUV spectrum (Fig.~\ref{fig1}b) used in the calculation features a strong and well-localized spectral phase variation at its center that may mimic the effect of a complex high-harmonic generation process or the resonant ionization phase of the target. As can be seen in Fig.~\ref{fig1}c,e, RABBITT is blind to the sharp phase variation, while the 1-2 quantum beat is particularly sensitive to it (Fig.~\ref{fig1}d,f).
The retrieved phase differences provide detailed information on the spectral phase $\varphi (E)$ well beyond its first derivative at the center, as we will show below. 
\begin{figure*}[t]
\includegraphics[width=\textwidth]{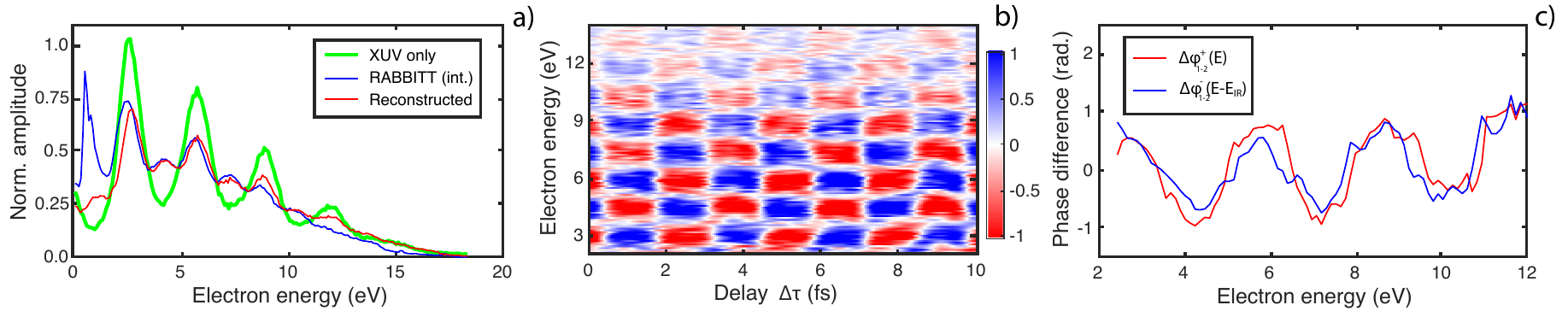}
\caption{\label{fig2} (a) XUV-only (green) and delay-integrated RABBITT spectrum (blue) from the experiment. The red curve results from the fit of the transition rates (Eq. \ref{Eq:8}) to the integrated RABBITT spectrum in the range from 3 eV to 12 eV. (b) Measured asymmetry signal, defined as in eq. \eqref{Eq:4}, as a function of pump-probe delay. (c). Retrieved phases $\Delta\varphi_{1-2}^+(E)$ and $\Delta\varphi_{1-2}^-(E-\hbar\omega_{\textsc{IR}})$ from the experiment.}
\end{figure*}

The XUV-APT spectrum is composed of odd high harmonics of an IR laser field, which result in mainbands (MBs) (one-photon-transitions) in the photoelectron spectrum, separated by twice the IR photon energy $\hbar \omega_{\textsc{IR}}$. Interaction with the IR probe leads to the appearance of sidebands (SB) between the mainbands, whose intensities oscillate as a function of the pump-probe delay $\tau$ at twice the IR laser frequency $2\omega_{\textsc{IR}}$ (Fig.~\ref{fig1}c). The beating is symmetric along the common light polarization axis as only partial waves with the same parity interfere. The phase offset of each sideband corresponds to the phase difference between the neighboring harmonics extracted by RABBITT $\Delta \varphi_{2-2}=\varphi(E+\hbar\omega_{\textsc{IR}})-\varphi(E-\hbar\omega_{\textsc{IR}})$. Since this phase difference can only be sampled at the sideband positions, the sharply structured phase profile remains undetected (Fig. \ref{fig1}e), even though the XUV spectrum spans the entire energy region.

In the 1-2 quantum beat method, by contrast, the interference of partial waves with opposite parity ($s-p$ or $p-d$) gives rise to an asymmetry of the electron angular distribution that beats at the angular frequency $\omega_{\textsc{IR}}$ as a function of $\tau$ (Fig. \ref{fig1} d) \cite{Laurent2012,Laurent2013}. This asymmetry, determined here by the difference of electron yield emitted to opposite sides of the plane perpendicular to the light polarization, is shown in Fig.~\ref{fig1} d. 
For the ultrashort APT employed here, both two-photon pathways (absorption and stimulated emission of an IR photon) can interfere with the one-photon amplitude across the whole spectral width of the APT. As long as the harmonics are spectrally sufficiently broad, therefore, the two phase differences $\Delta\varphi_{1-2}^+(E)=\varphi(E-\hbar\omega_{\textsc{IR}})-\varphi(E)$ and $\Delta\varphi_{1-2}^-(E)=\varphi(E)-\varphi(E+\hbar\omega_{\textsc{IR}})$ can both be retrieved continuously as a function of energy and the sharp phase profile is detected. As we will show below, the two phase differences can be retrieved fully analytically from the asymmetry trace. Furthermore, as the one-photon pathway is itself part of the interference, the retrieved phases are unaffected by the finite spectral bandwidth of the IR \cite{Busto2018}. For sharp resonances (see, e.g., Fig. \ref{fig1}), the 2-photon pathways serve as flat reference, such that the retrieved phases remain sharp.

The angle-dependent ionization probability is~\cite{Cirelli2018a}
\begin{equation}
I(E,\vartheta,\tau)=\Big|{\sum}_\ell(A^{+}_\ell+A^{-}_\ell)Y^0_\ell(\vartheta)
+iA^{1}_1 Y^0_1(\vartheta)\Big|^2,
\end{equation}
where $A^{1}_1$ and $A^{\pm}_\ell$ are the one-photon and two-photon amplitudes ($+/-$ designates \textsc{IR} absorption / emission, and $\ell=0,2$ is the photoelectron orbital angular momentum), $Y_{\ell}^m$ are spherical harmonics with $m=0$ due to the collinear alignment of the employed light fields, and $\vartheta$ is the angle between the electron photoemission direction and the common light polarization axis. The one- and two-photon amplitudes of the quantum pathways are functions of the kinetic energy $E$ and of the pump-probe delay $\tau$~\cite{Dahlstrom2013},
\begin{equation}
A^{1}_1=|A^{1}_1|e^{i\varphi(E)},\quad
A^{\pm}_\ell=|A^{\pm}_\ell|e^{i(\varphi^{\pm}_\ell(E)\pm\omega\tau)}.
\label{Eq:3}
\end{equation}
The spectral phase of the one-photon XUV ionization $\varphi(E) = \varphi_{\ell=1}^1(E)$ contains the EWS scattering phase and the XUV phase. The photoelectron asymmetry signal 
\begin{equation}
f_{a}(E,\tau) =  I(E,\tau)_{\vartheta \leq 90^{\circ}} - I(E,\tau)_{\vartheta \geq 90^{\circ}}, \label{Eq:4}
\end{equation}
given by the difference between emission into the forward and backward hemispheres, then follows as
\begin{equation}
\begin{split}
    f_{a}(E,\tau)={\sum}_{\sigma, \ell} &\sigma  c_{\ell} \left|A_1^1 \right| \left| A_{\ell}^{\sigma} \right| \\
    & \sin\left[\omega\tau+ \sigma \left( \varphi_{\ell}^{\sigma} (E) - \varphi_{1}^1 (E)   \right) \right], ~\label{Eq:5}
\end{split}
\end{equation}
where $\sigma=\pm$, $c_0=\sqrt{3}$ and $c_2=\sqrt{15}/4$.
For a simplified analytic estimate, the two-photon pathways can be approximated by the one-photon phase as $\varphi_{\ell}^{\sigma} (E) \simeq \varphi_1^1 (E-\sigma \hbar \omega_{\textsc{IR}} )$ since the method is sensitive only to phase variations but not to absolute phases. The cc-phase for different angular momenta~\cite{Fuchs2020} can be neglected since its variation is small within the present energy range. Likewise, the cc-transition probabilities to different $\ell$ are only weakly energy and $\ell$ dependent, such that $|A^{\sigma}_{\ell}|\approx |A^{\pm}|$~\cite{Busto2019}. \\
Consequently,
\begin{equation}
    f_a(E,\tau) \simeq A(E) \sin \left[\omega \tau + \delta(E) \right],\label{Eq:5a}
\end{equation}
where $A(E)$ and $\delta(E)$ are the modulus and phase of $a^+ e^{i\Delta\varphi^+_{1-2}}-{a^-}e^{i\Delta\varphi^-_{1-2}}$, with $a^{\sigma} (E)=\left| A_1^1 \right| \left| A^{\sigma} \right| \left( c_0 + c_2 \right) $ and $\Delta\varphi_{1-2}^{\sigma}(E)=\sigma ( \varphi(E - \sigma \hbar\omega_{\textsc{IR}})-\varphi(E))$.
This approximate relationship (Eq.~\eqref{Eq:5a}) illustrates the sensitivity of the 1-2 quantum beat method to rapid variations of the spectral phase. For energy-independent phases, $A(E)$ vanishes, as $a^{+}(E)\approx a^{-}(E)$. By contrast, phase differences $\Delta\varphi^{\pm}$ that vary rapidly within $\hbar\omega_{IR}$ result in strong oscillations of the photoemission asymmetry.
\begin{figure*}[t]
\includegraphics[width=\textwidth]{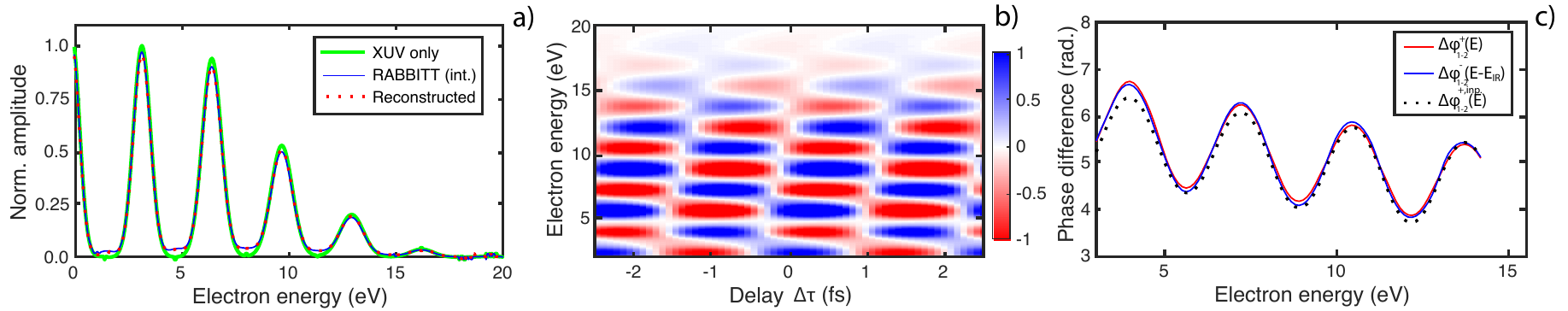}
\caption{\label{fig3} (a) XUV-only (green) and delay-integrated RABBITT spectrum (blue) from the quantum simulation employing a single-active electron (SAE) approximation and a model potential from \cite{Tong2005}. The red curve results from a fit of the transition rates (Eq. \ref{Eq:8}) to the integrated RABBITT spectrum in the range from 2 eV to 14 eV. (b) Calculated asymmetry signal, defined as in eq. \eqref{Eq:4}, as a function of pump-probe delay. (c) Retrieved phases $\Delta\varphi_{1-2}^+(E)$ (red) and $\Delta\varphi_{1-2}^-(E-\hbar\omega_{\textsc{IR}})$ (blue) in comparison to $\Delta\varphi_{1-2}^{+,inp.}(E)$ from the input phase.}
\end{figure*}

Fig. \ref{fig2} shows the results of a proof-of-principle experiment performed with atomic helium. The experiment is carried out resembling the RABBITT protocol and using an XUV-APT with spectrally broad high harmonics. The XUV-APT is generated via HHG using a $10$ fs FWHM IR laser pulse centered around $785$ nm from a carrier-envelope-phase (CEP) stabilized Ti:sapphire laser system. The CEP stabilization is essential for the observation of the asymmetry signal. Otherwise, an asymmetry would not be observable at all. The XUV-APT is focused together with a collinear time-delayed replica of the generating IR pulse on a cold helium gas jet. The resulting photoelectrons are collected with a cold target recoil ion momentum spectrometer (COLTRIMS) \cite{Dorner2000}, which allows for an angular resolved detection \cite{Heuser2016}. The setup is described in details in \cite{Sabbar2014}.
In the delay-integrated RABBITT spectrum, the MBs are depleted as compared to the XUV-only spectrum due to the IR induced cc-transitions to the SBs (Fig.~\ref{fig2}a). 
The asymmetry shows a checkerboard pattern (Fig.~\ref{fig2}b), which implies a characteristic energy dependence of the spectral phase. If the phase were spectrally flat, only a weak and constant asymmetry signal comparable to the upper (or lower) part in Fig.~\ref{fig1}d would be expected.  A similar checkerboard has been observed in recent experiments exploiting the 1-2-quantum beat, where both even and odd harmonics~\cite{Laurent2012, Laurent2013} have been employed, revealing a non-flat phase behavior, as well.

The retrieval of the phase differences $\Delta \varphi_{1-2}^{\pm}$ from the asymmetry comprises three steps. First, we determine the modulus of the one-photon amplitude $|A^{1}_1(E)|=\sqrt{f_{tot}^{XUV}(E)}$ from an XUV-only spectrum.
Second, the modulus of the two-photon amplitudes for absorption and emission $|A^{\pm}|$ are determined by fitting the IR-transitions rates to the delay-integrated RABBITT spectrum (Fig.~\ref{fig2}a). The amplitudes of the two-photon pathways are replicas of the one-photon amplitudes, shifted by the IR photon energy:
\begin{align}
A^{+}(E)&=r^+(E)A^{1}_1(E-\hbar\omega_{\textsc{IR}})\label{Eq:7}\\
A^{-}(E)&=r^-(E)A^{1}_1(E+\hbar\omega_{\textsc{IR}}),
\label{Eq:8}
\end{align}
with $r^{\pm}(E)=c^{\pm}+d^{\pm}E$. The parameters $c^{\pm}$ and $d^{\pm}$ account for a smooth energy dependence of the cc-transition rates and are fitted to the delay-integrated RABBITT spectrum (see SM),
\begin{equation}
\label{Eq:9}
\langle f_{tot}(E,\tau)\rangle_{\tau}=|A^{1}(E)|^2+2|A^{+}(E)|^2+2|A^{-}(E)|^2.
\end{equation}
Finally, using $a^+(E)$ and $a^-(E)$,  we can analytically determine   $\Delta\varphi^\pm_{1-2}(E)$ from the measured amplitude $A(E)$ and phase $\delta(E)$ of the asymmetry signal as continuous function of the energy via Eq.~\eqref{Eq:5a} (see SM). We note that for ionization from other than $s$-ground states, the parametrization of the angular dependent ionization amplitude must be extended to account for partial waves with different l- and m-quantum numbers. For the procedure to be consistent, the retrieved phase differences must satisfy the identity $\Delta\varphi_{1-2}^+(E)=\Delta\varphi_{1-2}^-(E-\hbar\omega_{\textsc{IR}})$. Figure \ref{fig2}c shows that $\Delta\varphi_{1-2}^+(E)$ and $\Delta\varphi_{1-2}^-(E-\hbar\omega_{\textsc{IR}})$ are indeed in close agreement with each other across a wide energy range, demonstrating the applicability of the phase retrieval. The deviation of the two phases for energies slightly above 6 eV and 9 eV indicates a larger uncertainty for these energies.

The retrieved phase differences from the 1-2 quantum beat method exhibit periodic oscillations with the same periodicity as the XUV harmonics. Since in this energy region neither the EWS scattering phase nor the cc-phase of atomic helium oscillates, the phase oscillations can be attributed to the ionizing APT. 
To support this hypothesis, we simulate the experiment by solving the TDSE in the single-active-electron (SAE) approximation \cite{Douguet2016}. We have checked for several delay steps that a full 2-electron calculation \cite{Feist2008a, Donsa2019b} yields indistinguishable results. As input we use an XUV pulse featuring spectral phase oscillations. As expected, we obtain an asymmetry signal exhibiting a qualitatively similar checkerboard pattern (Fig. \ref{fig3}b). The tilt observed in the pattern is due to the attochirp of the pulse.
We further verify the retrieval method by applying it to the simulated data and comparing the result to the original XUV phase. Figure \ref{fig3}c shows the excellent agreement between the phase difference $\Delta\varphi_{1-2}^{+,inp.}$ from the input phase and the retrieved phase differences $\Delta\varphi^+_{1-2}$ and $\Delta\varphi^-_{1-2}$. The small deviation of the latter two across the full energy range indicates the accurate phase retrieval for all energies. The slight deviation with respect to the input phase at low kinetic energies is due to the EWS and IR-induced cc-phase (see approximations in eq. \eqref{Eq:5a}), which are no longer negligible at these energies and cannot be separated from the XUV phase by the retrieval method.

The 1-2 quantum beat method enables the measurement of phase variations across an individual harmonic. This is fundamentally different from measuring phase differences between the same spectral region of different harmonics. Therefore, this method gives us the unprecedented ability to simultaneously measure the atto- and the femtochirp of the APT inherent to the HHG process~\cite{Varju2005}. The attochirp, which corresponds to a linear increase (or decrease) of the group delay across the full spectrum, is encoded in the slope of the mean of $\Delta \varphi_{1-2}^{\pm}(E)$. In the time domain, the attochirp translates into different harmonics being emitted at different times during the IR cycle, stretching each attosecond burst. 

\begin{figure}[t]
\includegraphics[width=0.48\textwidth]{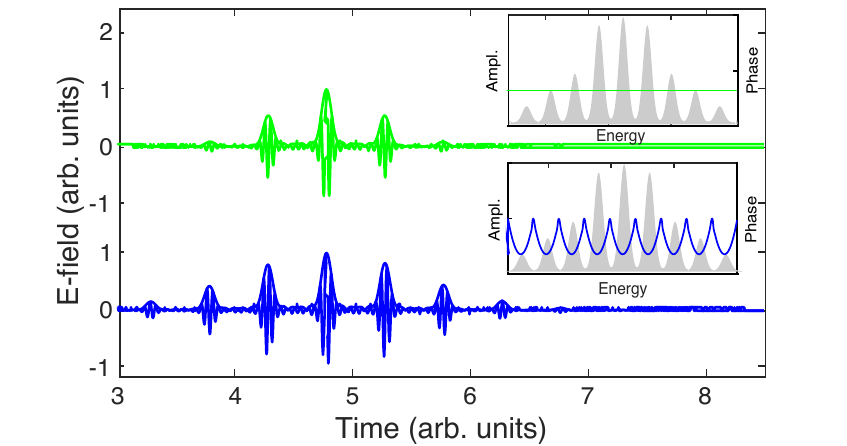}
\caption{\label{fig4} Comparison of an attosecond pulse train (APT) with periodic oscillations of the spectral phase (blue) and an APT with a flat phase (green). The spectrum is identical for both APTs. The insets show the spectrum and the corresponding phase for both APTs.}
\end{figure}

The femtochirp corresponds to the observed oscillations of $\Delta\varphi_{1-2}^{\pm}(E)$.
Such a rapidly varying phase within a given harmonic in the plateau region was originally predicted more than 15 years ago~\cite{Varju2005}. 
The femtochirp results from the interplay of two microscopic effects.
First, the phase of each harmonic depends on the IR intensity at the time of tunnel ionization~\cite{Lewenstein1995}. The use of ultrashort pulsed light sources to drive HHG implies a rapidly varying intensity envelope, which results in fine-scale phase structures within each harmonic.
Second, several quantum paths contribute, in general, to the generation of the harmonics in the plateau region. Even though the intensity dependence of the phase for each path is approximately linear, the superposition of multiple paths with different phase drifts leads to a complex phase structure within each harmonic~\cite{Gaarde2002}.
Both effects, therefore, can give rise to a femtochirp, which, in the time domain, results in an unequal spacing of the attosecond bursts \cite{Varju2005} and stretches the envelope of the pulse train, see Fig. \ref{fig4}.
As the multi-quantum-path interference is sensitive to and easily suppressed by macroscopic propagation effects, we expect the intensity envelope effect to be the dominant contribution under realistic experimental conditions.

In conclusion, we have shown that the 1-2 quantum beat method can be used to retrieve phase variations of a photoelectron wave packet as a continuous function of energy, with a finer energy resolution than the probe frequency spectral width. In particular, we demonstrate with a proof-of-principle experiment that the 1-2-quantum beat method allows us to observe the strong periodic modulations of the spectral phase due to the harmonic chirp and caused by the HHG process itself. Despite the harmonic chirp being an already well-established concept~\cite{Shin1999, Shin2001, Mairesse2005, Haessler2012, Ardana-Lamas2016, Kim2013}, such phase modulations could not have been observed so far, since they are inherently inaccessible to any pump-probe scheme that relies on the comparison of the phase of consecutive harmonics such as RABBITT or related techniques~\cite{Laurent2012, Loriot2017, Donsa2019a, You2020}. As the retrieval method returns phase differences as a continuous function of energy and is given in closed form, it constitutes a valuable tool to investigate even more complex photoionization dynamics and provides unprecedented access to the spectral phase of wave packets resulting from the break-up of quantum systems.

\begin{acknowledgments}
All authors acknowledge the COST Action CA18222 (Attosecond Chemistry).
J.F., L.C. and U.K. acknowledge the support of the NCCR MUST, funded by the Swiss National Science Foundation.
L.A. and N.D. acknowledge the support of the United States National Science Foundation under NSF Grant No. PHY-1607588 and by the UCF in-house OR Grant Acc. No. 24089045.
Parts of the calculations were performed on the Vienna Scientific Cluster (VSC3). S.D. and J.B. acknowledge the support by the WWTF through Project No. MA14-002, and the FWF through Projects No. FWF-SFB041-VICOM, and No. FWF-W1243-Solids4Fun, as well as the IMPRS-APS. 
F.M. acknowledges the MICINN projects PID2019-105458RB-I00, the 'Severo Ochoa' Programme for Centres of Excellence in R{\&}D (SEV-2016-0686) and the 'María de Maeztu' Programme for Units of Excellence in R{\&}D (CEX2018-000805-M).
\end{acknowledgments}

\bibliography{Phd_Paper_1w}

\begin{thebibliography}{54}%
\makeatletter
\providecommand \@ifxundefined [1]{%
 \@ifx{#1\undefined}
}%
\providecommand \@ifnum [1]{%
 \ifnum #1\expandafter \@firstoftwo
 \else \expandafter \@secondoftwo
 \fi
}%
\providecommand \@ifx [1]{%
 \ifx #1\expandafter \@firstoftwo
 \else \expandafter \@secondoftwo
 \fi
}%
\providecommand \natexlab [1]{#1}%
\providecommand \enquote  [1]{``#1''}%
\providecommand \bibnamefont  [1]{#1}%
\providecommand \bibfnamefont [1]{#1}%
\providecommand \citenamefont [1]{#1}%
\providecommand \href@noop [0]{\@secondoftwo}%
\providecommand \href [0]{\begingroup \@sanitize@url \@href}%
\providecommand \@href[1]{\@@startlink{#1}\@@href}%
\providecommand \@@href[1]{\endgroup#1\@@endlink}%
\providecommand \@sanitize@url [0]{\catcode `\\12\catcode `\$12\catcode
  `\&12\catcode `\#12\catcode `\^12\catcode `\_12\catcode `\%12\relax}%
\providecommand \@@startlink[1]{}%
\providecommand \@@endlink[0]{}%
\providecommand \url  [0]{\begingroup\@sanitize@url \@url }%
\providecommand \@url [1]{\endgroup\@href {#1}{\urlprefix }}%
\providecommand \urlprefix  [0]{URL }%
\providecommand \Eprint [0]{\href }%
\providecommand \doibase [0]{http://dx.doi.org/}%
\providecommand \selectlanguage [0]{\@gobble}%
\providecommand \bibinfo  [0]{\@secondoftwo}%
\providecommand \bibfield  [0]{\@secondoftwo}%
\providecommand \translation [1]{[#1]}%
\providecommand \BibitemOpen [0]{}%
\providecommand \bibitemStop [0]{}%
\providecommand \bibitemNoStop [0]{.\EOS\space}%
\providecommand \EOS [0]{\spacefactor3000\relax}%
\providecommand \BibitemShut  [1]{\csname bibitem#1\endcsname}%
\let\auto@bib@innerbib\@empty
\bibitem [{\citenamefont {Schultze}\ \emph {et~al.}(2010)\citenamefont
  {Schultze}, \citenamefont {Fiess}, \citenamefont {Karpowicz}, \citenamefont
  {Gagnon}, \citenamefont {Korbman}, \citenamefont {Hofstetter}, \citenamefont
  {Neppl}, \citenamefont {Cavalieri}, \citenamefont {Komninos}, \citenamefont
  {Mercouris}, \citenamefont {Nicolaides}, \citenamefont {Pazourek},
  \citenamefont {Nagele}, \citenamefont {Feist}, \citenamefont
  {Burgd{\"{o}}rfer}, \citenamefont {Azzeer}, \citenamefont {Ernstorfer},
  \citenamefont {Kienberger}, \citenamefont {Kleineberg}, \citenamefont
  {Goulielmakis}, \citenamefont {Krausz},\ and\ \citenamefont
  {Yakovlev}}]{Schultze2010}%
  \BibitemOpen
  \bibfield  {author} {\bibinfo {author} {\bibfnamefont {M.}~\bibnamefont
  {Schultze}}, \bibinfo {author} {\bibfnamefont {M.}~\bibnamefont {Fiess}},
  \bibinfo {author} {\bibfnamefont {N.}~\bibnamefont {Karpowicz}}, \bibinfo
  {author} {\bibfnamefont {J.}~\bibnamefont {Gagnon}}, \bibinfo {author}
  {\bibfnamefont {M.}~\bibnamefont {Korbman}}, \bibinfo {author} {\bibfnamefont
  {M.}~\bibnamefont {Hofstetter}}, \bibinfo {author} {\bibfnamefont
  {S.}~\bibnamefont {Neppl}}, \bibinfo {author} {\bibfnamefont {A.~L.}\
  \bibnamefont {Cavalieri}}, \bibinfo {author} {\bibfnamefont {Y.}~\bibnamefont
  {Komninos}}, \bibinfo {author} {\bibfnamefont {T.}~\bibnamefont {Mercouris}},
  \bibinfo {author} {\bibfnamefont {C.~A.}\ \bibnamefont {Nicolaides}},
  \bibinfo {author} {\bibfnamefont {R.}~\bibnamefont {Pazourek}}, \bibinfo
  {author} {\bibfnamefont {S.}~\bibnamefont {Nagele}}, \bibinfo {author}
  {\bibfnamefont {J.}~\bibnamefont {Feist}}, \bibinfo {author} {\bibfnamefont
  {J.}~\bibnamefont {Burgd{\"{o}}rfer}}, \bibinfo {author} {\bibfnamefont
  {A.~M.}\ \bibnamefont {Azzeer}}, \bibinfo {author} {\bibfnamefont
  {R.}~\bibnamefont {Ernstorfer}}, \bibinfo {author} {\bibfnamefont
  {R.}~\bibnamefont {Kienberger}}, \bibinfo {author} {\bibfnamefont
  {U.}~\bibnamefont {Kleineberg}}, \bibinfo {author} {\bibfnamefont
  {E.}~\bibnamefont {Goulielmakis}}, \bibinfo {author} {\bibfnamefont
  {F.}~\bibnamefont {Krausz}}, \ and\ \bibinfo {author} {\bibfnamefont {V.~S.}\
  \bibnamefont {Yakovlev}},\ }\href {\doibase 10.1126/science.1189401}
  {\bibfield  {journal} {\bibinfo  {journal} {Science}\ }\textbf {\bibinfo
  {volume} {328}},\ \bibinfo {pages} {1658} (\bibinfo {year}
  {2010})}\BibitemShut {NoStop}%
\bibitem [{\citenamefont {Isinger}\ \emph {et~al.}(2017)\citenamefont
  {Isinger}, \citenamefont {Squibb}, \citenamefont {Busto}, \citenamefont
  {Zhong}, \citenamefont {Harth}, \citenamefont {Kroon}, \citenamefont {Nandi},
  \citenamefont {Arnold}, \citenamefont {Miranda}, \citenamefont
  {Dahlstr{\"{o}}m}, \citenamefont {Lindroth}, \citenamefont {Feifel},
  \citenamefont {Gisselbrecht},\ and\ \citenamefont
  {L'Huillier}}]{Isinger2017}%
  \BibitemOpen
  \bibfield  {author} {\bibinfo {author} {\bibfnamefont {M.}~\bibnamefont
  {Isinger}}, \bibinfo {author} {\bibfnamefont {R.~J.}\ \bibnamefont {Squibb}},
  \bibinfo {author} {\bibfnamefont {D.}~\bibnamefont {Busto}}, \bibinfo
  {author} {\bibfnamefont {S.}~\bibnamefont {Zhong}}, \bibinfo {author}
  {\bibfnamefont {A.}~\bibnamefont {Harth}}, \bibinfo {author} {\bibfnamefont
  {D.}~\bibnamefont {Kroon}}, \bibinfo {author} {\bibfnamefont
  {S.}~\bibnamefont {Nandi}}, \bibinfo {author} {\bibfnamefont {C.~L.}\
  \bibnamefont {Arnold}}, \bibinfo {author} {\bibfnamefont {M.}~\bibnamefont
  {Miranda}}, \bibinfo {author} {\bibfnamefont {J.~M.}\ \bibnamefont
  {Dahlstr{\"{o}}m}}, \bibinfo {author} {\bibfnamefont {E.}~\bibnamefont
  {Lindroth}}, \bibinfo {author} {\bibfnamefont {R.}~\bibnamefont {Feifel}},
  \bibinfo {author} {\bibfnamefont {M.}~\bibnamefont {Gisselbrecht}}, \ and\
  \bibinfo {author} {\bibfnamefont {A.}~\bibnamefont {L'Huillier}},\ }\href
  {\doibase 10.1126/science.aao7043} {\bibfield  {journal} {\bibinfo  {journal}
  {Science}\ }\textbf {\bibinfo {volume} {358}},\ \bibinfo {pages} {893}
  (\bibinfo {year} {2017})},\ \Eprint {http://arxiv.org/abs/1709.01780}
  {arXiv:1709.01780} \BibitemShut {NoStop}%
\bibitem [{\citenamefont {Cirelli}\ \emph {et~al.}(2018)\citenamefont
  {Cirelli}, \citenamefont {Marante}, \citenamefont {Heuser}, \citenamefont
  {Petersson}, \citenamefont {Gal{\'{a}}n}, \citenamefont {Argenti},
  \citenamefont {Zhong}, \citenamefont {Busto}, \citenamefont {Isinger},
  \citenamefont {Nandi}, \citenamefont {MacLot}, \citenamefont {Rading},
  \citenamefont {Johnsson}, \citenamefont {Gisselbrecht}, \citenamefont
  {Lucchini}, \citenamefont {Gallmann}, \citenamefont {Dahlstr{\"{o}}m},
  \citenamefont {Lindroth}, \citenamefont {L'Huillier}, \citenamefont
  {Mart{\'{i}}n},\ and\ \citenamefont {Keller}}]{Cirelli2018a}%
  \BibitemOpen
  \bibfield  {author} {\bibinfo {author} {\bibfnamefont {C.}~\bibnamefont
  {Cirelli}}, \bibinfo {author} {\bibfnamefont {C.}~\bibnamefont {Marante}},
  \bibinfo {author} {\bibfnamefont {S.}~\bibnamefont {Heuser}}, \bibinfo
  {author} {\bibfnamefont {C.~L.}\ \bibnamefont {Petersson}}, \bibinfo {author}
  {\bibfnamefont {{\'{A}}.~J.}\ \bibnamefont {Gal{\'{a}}n}}, \bibinfo {author}
  {\bibfnamefont {L.}~\bibnamefont {Argenti}}, \bibinfo {author} {\bibfnamefont
  {S.}~\bibnamefont {Zhong}}, \bibinfo {author} {\bibfnamefont
  {D.}~\bibnamefont {Busto}}, \bibinfo {author} {\bibfnamefont
  {M.}~\bibnamefont {Isinger}}, \bibinfo {author} {\bibfnamefont
  {S.}~\bibnamefont {Nandi}}, \bibinfo {author} {\bibfnamefont
  {S.}~\bibnamefont {MacLot}}, \bibinfo {author} {\bibfnamefont
  {L.}~\bibnamefont {Rading}}, \bibinfo {author} {\bibfnamefont
  {P.}~\bibnamefont {Johnsson}}, \bibinfo {author} {\bibfnamefont
  {M.}~\bibnamefont {Gisselbrecht}}, \bibinfo {author} {\bibfnamefont
  {M.}~\bibnamefont {Lucchini}}, \bibinfo {author} {\bibfnamefont
  {L.}~\bibnamefont {Gallmann}}, \bibinfo {author} {\bibfnamefont {J.~M.}\
  \bibnamefont {Dahlstr{\"{o}}m}}, \bibinfo {author} {\bibfnamefont
  {E.}~\bibnamefont {Lindroth}}, \bibinfo {author} {\bibfnamefont
  {A.}~\bibnamefont {L'Huillier}}, \bibinfo {author} {\bibfnamefont
  {F.}~\bibnamefont {Mart{\'{i}}n}}, \ and\ \bibinfo {author} {\bibfnamefont
  {U.}~\bibnamefont {Keller}},\ }\href {\doibase 10.1038/s41467-018-03009-1}
  {\bibfield  {journal} {\bibinfo  {journal} {Nat. Commun.}\ }\textbf {\bibinfo
  {volume} {9}},\ \bibinfo {pages} {955} (\bibinfo {year} {2018})}\BibitemShut
  {NoStop}%
\bibitem [{\citenamefont {Sansone}\ \emph {et~al.}(2010)\citenamefont
  {Sansone}, \citenamefont {Kelkensberg}, \citenamefont {P{\'{e}}rez-Torres},
  \citenamefont {Morales}, \citenamefont {Kling}, \citenamefont {Siu},
  \citenamefont {Ghafur}, \citenamefont {Johnsson}, \citenamefont {Swoboda},
  \citenamefont {Benedetti}, \citenamefont {Ferrari}, \citenamefont
  {L{\'{e}}pine}, \citenamefont {Sanz-Vicario}, \citenamefont {Zherebtsov},
  \citenamefont {Znakovskaya}, \citenamefont {Lhuillier}, \citenamefont
  {Ivanov}, \citenamefont {Nisoli}, \citenamefont {Mart{\'{i}}n},\ and\
  \citenamefont {Vrakking}}]{Sansone2010}%
  \BibitemOpen
  \bibfield  {author} {\bibinfo {author} {\bibfnamefont {G.}~\bibnamefont
  {Sansone}}, \bibinfo {author} {\bibfnamefont {F.}~\bibnamefont
  {Kelkensberg}}, \bibinfo {author} {\bibfnamefont {J.~F.}\ \bibnamefont
  {P{\'{e}}rez-Torres}}, \bibinfo {author} {\bibfnamefont {F.}~\bibnamefont
  {Morales}}, \bibinfo {author} {\bibfnamefont {M.~F.}\ \bibnamefont {Kling}},
  \bibinfo {author} {\bibfnamefont {W.}~\bibnamefont {Siu}}, \bibinfo {author}
  {\bibfnamefont {O.}~\bibnamefont {Ghafur}}, \bibinfo {author} {\bibfnamefont
  {P.}~\bibnamefont {Johnsson}}, \bibinfo {author} {\bibfnamefont
  {M.}~\bibnamefont {Swoboda}}, \bibinfo {author} {\bibfnamefont
  {E.}~\bibnamefont {Benedetti}}, \bibinfo {author} {\bibfnamefont
  {F.}~\bibnamefont {Ferrari}}, \bibinfo {author} {\bibfnamefont
  {F.}~\bibnamefont {L{\'{e}}pine}}, \bibinfo {author} {\bibfnamefont {J.~L.}\
  \bibnamefont {Sanz-Vicario}}, \bibinfo {author} {\bibfnamefont
  {S.}~\bibnamefont {Zherebtsov}}, \bibinfo {author} {\bibfnamefont
  {I.}~\bibnamefont {Znakovskaya}}, \bibinfo {author} {\bibfnamefont
  {A.}~\bibnamefont {Lhuillier}}, \bibinfo {author} {\bibfnamefont {M.~Y.}\
  \bibnamefont {Ivanov}}, \bibinfo {author} {\bibfnamefont {M.}~\bibnamefont
  {Nisoli}}, \bibinfo {author} {\bibfnamefont {F.}~\bibnamefont
  {Mart{\'{i}}n}}, \ and\ \bibinfo {author} {\bibfnamefont {M.~J.}\
  \bibnamefont {Vrakking}},\ }\href {\doibase 10.1038/nature09084} {\bibfield
  {journal} {\bibinfo  {journal} {Nature}\ }\textbf {\bibinfo {volume} {465}},\
  \bibinfo {pages} {763} (\bibinfo {year} {2010})}\BibitemShut {NoStop}%
\bibitem [{\citenamefont {Vos}\ \emph {et~al.}(2018)\citenamefont {Vos},
  \citenamefont {Cattaneo}, \citenamefont {Patchkovskii}, \citenamefont
  {Zimmermann}, \citenamefont {Cirelli}, \citenamefont {Lucchini},
  \citenamefont {Kheifets}, \citenamefont {Landsman},\ and\ \citenamefont
  {Keller}}]{Vos2018}%
  \BibitemOpen
  \bibfield  {author} {\bibinfo {author} {\bibfnamefont {J.}~\bibnamefont
  {Vos}}, \bibinfo {author} {\bibfnamefont {L.}~\bibnamefont {Cattaneo}},
  \bibinfo {author} {\bibfnamefont {S.}~\bibnamefont {Patchkovskii}}, \bibinfo
  {author} {\bibfnamefont {T.}~\bibnamefont {Zimmermann}}, \bibinfo {author}
  {\bibfnamefont {C.}~\bibnamefont {Cirelli}}, \bibinfo {author} {\bibfnamefont
  {M.}~\bibnamefont {Lucchini}}, \bibinfo {author} {\bibfnamefont
  {A.}~\bibnamefont {Kheifets}}, \bibinfo {author} {\bibfnamefont {A.~S.}\
  \bibnamefont {Landsman}}, \ and\ \bibinfo {author} {\bibfnamefont
  {U.}~\bibnamefont {Keller}},\ }\href {\doibase 10.1126/science.aao4731}
  {\bibfield  {journal} {\bibinfo  {journal} {Science}\ }\textbf {\bibinfo
  {volume} {360}},\ \bibinfo {pages} {1326} (\bibinfo {year}
  {2018})}\BibitemShut {NoStop}%
\bibitem [{\citenamefont {Cavalieri}\ \emph {et~al.}(2007)\citenamefont
  {Cavalieri}, \citenamefont {M{\"{u}}ller}, \citenamefont {Uphues},
  \citenamefont {Yakovlev}, \citenamefont {Baltu{\v{s}}ka}, \citenamefont
  {Horvath}, \citenamefont {Schmidt}, \citenamefont {Bl{\"{u}}mel},
  \citenamefont {Holzwarth}, \citenamefont {Hendel}, \citenamefont {Drescher},
  \citenamefont {Kleineberg}, \citenamefont {Echenique}, \citenamefont
  {Kienberger}, \citenamefont {Krausz},\ and\ \citenamefont
  {Heinzmann}}]{Cavalieri2007}%
  \BibitemOpen
  \bibfield  {author} {\bibinfo {author} {\bibfnamefont {A.~L.}\ \bibnamefont
  {Cavalieri}}, \bibinfo {author} {\bibfnamefont {N.}~\bibnamefont
  {M{\"{u}}ller}}, \bibinfo {author} {\bibfnamefont {T.}~\bibnamefont
  {Uphues}}, \bibinfo {author} {\bibfnamefont {V.~S.}\ \bibnamefont
  {Yakovlev}}, \bibinfo {author} {\bibfnamefont {A.}~\bibnamefont
  {Baltu{\v{s}}ka}}, \bibinfo {author} {\bibfnamefont {B.}~\bibnamefont
  {Horvath}}, \bibinfo {author} {\bibfnamefont {B.}~\bibnamefont {Schmidt}},
  \bibinfo {author} {\bibfnamefont {L.}~\bibnamefont {Bl{\"{u}}mel}}, \bibinfo
  {author} {\bibfnamefont {R.}~\bibnamefont {Holzwarth}}, \bibinfo {author}
  {\bibfnamefont {S.}~\bibnamefont {Hendel}}, \bibinfo {author} {\bibfnamefont
  {M.}~\bibnamefont {Drescher}}, \bibinfo {author} {\bibfnamefont
  {U.}~\bibnamefont {Kleineberg}}, \bibinfo {author} {\bibfnamefont {P.~M.}\
  \bibnamefont {Echenique}}, \bibinfo {author} {\bibfnamefont {R.}~\bibnamefont
  {Kienberger}}, \bibinfo {author} {\bibfnamefont {F.}~\bibnamefont {Krausz}},
  \ and\ \bibinfo {author} {\bibfnamefont {U.}~\bibnamefont {Heinzmann}},\
  }\href {\doibase 10.1038/nature06229} {\bibfield  {journal} {\bibinfo
  {journal} {Nature}\ }\textbf {\bibinfo {volume} {449}},\ \bibinfo {pages}
  {1029} (\bibinfo {year} {2007})}\BibitemShut {NoStop}%
\bibitem [{\citenamefont {Locher}\ \emph {et~al.}(2015)\citenamefont {Locher},
  \citenamefont {Castiglioni}, \citenamefont {Lucchini}, \citenamefont {Greif},
  \citenamefont {Gallmann}, \citenamefont {Osterwalder}, \citenamefont
  {Hengsberger},\ and\ \citenamefont {Keller}}]{Locher2015}%
  \BibitemOpen
  \bibfield  {author} {\bibinfo {author} {\bibfnamefont {R.}~\bibnamefont
  {Locher}}, \bibinfo {author} {\bibfnamefont {L.}~\bibnamefont {Castiglioni}},
  \bibinfo {author} {\bibfnamefont {M.}~\bibnamefont {Lucchini}}, \bibinfo
  {author} {\bibfnamefont {M.}~\bibnamefont {Greif}}, \bibinfo {author}
  {\bibfnamefont {L.}~\bibnamefont {Gallmann}}, \bibinfo {author}
  {\bibfnamefont {J.}~\bibnamefont {Osterwalder}}, \bibinfo {author}
  {\bibfnamefont {M.}~\bibnamefont {Hengsberger}}, \ and\ \bibinfo {author}
  {\bibfnamefont {U.}~\bibnamefont {Keller}},\ }\href {\doibase
  10.1364/OPTICA.2.000405} {\bibfield  {journal} {\bibinfo  {journal} {Optica}\
  }\textbf {\bibinfo {volume} {2}},\ \bibinfo {pages} {405} (\bibinfo {year}
  {2015})}\BibitemShut {NoStop}%
\bibitem [{\citenamefont {Kasmi}\ \emph {et~al.}(2017)\citenamefont {Kasmi},
  \citenamefont {Lucchini}, \citenamefont {Castiglioni}, \citenamefont
  {Kliuiev}, \citenamefont {Osterwalder}, \citenamefont {Hengsberger},
  \citenamefont {Gallmann}, \citenamefont {Kr{\"{u}}ger},\ and\ \citenamefont
  {Keller}}]{Kasmi2017}%
  \BibitemOpen
  \bibfield  {author} {\bibinfo {author} {\bibfnamefont {L.}~\bibnamefont
  {Kasmi}}, \bibinfo {author} {\bibfnamefont {M.}~\bibnamefont {Lucchini}},
  \bibinfo {author} {\bibfnamefont {L.}~\bibnamefont {Castiglioni}}, \bibinfo
  {author} {\bibfnamefont {P.}~\bibnamefont {Kliuiev}}, \bibinfo {author}
  {\bibfnamefont {J.}~\bibnamefont {Osterwalder}}, \bibinfo {author}
  {\bibfnamefont {M.}~\bibnamefont {Hengsberger}}, \bibinfo {author}
  {\bibfnamefont {L.}~\bibnamefont {Gallmann}}, \bibinfo {author}
  {\bibfnamefont {P.}~\bibnamefont {Kr{\"{u}}ger}}, \ and\ \bibinfo {author}
  {\bibfnamefont {U.}~\bibnamefont {Keller}},\ }\href {\doibase
  10.1364/OPTICA.4.001492} {\bibfield  {journal} {\bibinfo  {journal} {Optica}\
  }\textbf {\bibinfo {volume} {4}},\ \bibinfo {pages} {1492} (\bibinfo {year}
  {2017})}\BibitemShut {NoStop}%
\bibitem [{\citenamefont {Ossiander}\ \emph {et~al.}(2018)\citenamefont
  {Ossiander}, \citenamefont {Riemensberger}, \citenamefont {Neppl},
  \citenamefont {Mittermair}, \citenamefont {Sch{\"{a}}ffer}, \citenamefont
  {Duensing}, \citenamefont {Wagner}, \citenamefont {Heider}, \citenamefont
  {Wurzer}, \citenamefont {Gerl}, \citenamefont {Schnitzenbaumer},
  \citenamefont {Barth}, \citenamefont {Libisch}, \citenamefont {Lemell},
  \citenamefont {Burgd{\"{o}}rfer}, \citenamefont {Feulner},\ and\
  \citenamefont {Kienberger}}]{Ossiander2018}%
  \BibitemOpen
  \bibfield  {author} {\bibinfo {author} {\bibfnamefont {M.}~\bibnamefont
  {Ossiander}}, \bibinfo {author} {\bibfnamefont {J.}~\bibnamefont
  {Riemensberger}}, \bibinfo {author} {\bibfnamefont {S.}~\bibnamefont
  {Neppl}}, \bibinfo {author} {\bibfnamefont {M.}~\bibnamefont {Mittermair}},
  \bibinfo {author} {\bibfnamefont {M.}~\bibnamefont {Sch{\"{a}}ffer}},
  \bibinfo {author} {\bibfnamefont {A.}~\bibnamefont {Duensing}}, \bibinfo
  {author} {\bibfnamefont {M.~S.}\ \bibnamefont {Wagner}}, \bibinfo {author}
  {\bibfnamefont {R.}~\bibnamefont {Heider}}, \bibinfo {author} {\bibfnamefont
  {M.}~\bibnamefont {Wurzer}}, \bibinfo {author} {\bibfnamefont
  {M.}~\bibnamefont {Gerl}}, \bibinfo {author} {\bibfnamefont {M.}~\bibnamefont
  {Schnitzenbaumer}}, \bibinfo {author} {\bibfnamefont {J.~V.}\ \bibnamefont
  {Barth}}, \bibinfo {author} {\bibfnamefont {F.}~\bibnamefont {Libisch}},
  \bibinfo {author} {\bibfnamefont {C.}~\bibnamefont {Lemell}}, \bibinfo
  {author} {\bibfnamefont {J.}~\bibnamefont {Burgd{\"{o}}rfer}}, \bibinfo
  {author} {\bibfnamefont {P.}~\bibnamefont {Feulner}}, \ and\ \bibinfo
  {author} {\bibfnamefont {R.}~\bibnamefont {Kienberger}},\ }\href {\doibase
  10.1038/s41586-018-0503-6} {\enquote {\bibinfo {title} {{Absolute timing of
  the photoelectric effect}},}\ } (\bibinfo {year} {2018})\BibitemShut
  {NoStop}%
\bibitem [{\citenamefont {Pazourek}\ \emph {et~al.}(2012)\citenamefont
  {Pazourek}, \citenamefont {Feist}, \citenamefont {Nagele},\ and\
  \citenamefont {Burgd{\"{o}}rfer}}]{Pazourek2012}%
  \BibitemOpen
  \bibfield  {author} {\bibinfo {author} {\bibfnamefont {R.}~\bibnamefont
  {Pazourek}}, \bibinfo {author} {\bibfnamefont {J.}~\bibnamefont {Feist}},
  \bibinfo {author} {\bibfnamefont {S.}~\bibnamefont {Nagele}}, \ and\ \bibinfo
  {author} {\bibfnamefont {J.}~\bibnamefont {Burgd{\"{o}}rfer}},\ }\href
  {\doibase 10.1103/PhysRevLett.108.163001} {\bibfield  {journal} {\bibinfo
  {journal} {Phys. Rev. Lett.}\ }\textbf {\bibinfo {volume} {108}},\ \bibinfo
  {pages} {163001} (\bibinfo {year} {2012})},\ \Eprint
  {http://arxiv.org/abs/1112.3273} {arXiv:1112.3273} \BibitemShut {NoStop}%
\bibitem [{\citenamefont {Marante}\ \emph {et~al.}(2017)\citenamefont
  {Marante}, \citenamefont {Klinker}, \citenamefont {Kjellsson}, \citenamefont
  {Lindroth}, \citenamefont {Gonz{\'{a}}lez-V{\'{a}}zquez}, \citenamefont
  {Argenti},\ and\ \citenamefont {Mart{\'{i}}n}}]{Marante2017}%
  \BibitemOpen
  \bibfield  {author} {\bibinfo {author} {\bibfnamefont {C.}~\bibnamefont
  {Marante}}, \bibinfo {author} {\bibfnamefont {M.}~\bibnamefont {Klinker}},
  \bibinfo {author} {\bibfnamefont {T.}~\bibnamefont {Kjellsson}}, \bibinfo
  {author} {\bibfnamefont {E.}~\bibnamefont {Lindroth}}, \bibinfo {author}
  {\bibfnamefont {J.}~\bibnamefont {Gonz{\'{a}}lez-V{\'{a}}zquez}}, \bibinfo
  {author} {\bibfnamefont {L.}~\bibnamefont {Argenti}}, \ and\ \bibinfo
  {author} {\bibfnamefont {F.}~\bibnamefont {Mart{\'{i}}n}},\ }\href {\doibase
  10.1103/PhysRevA.96.022507} {\bibfield  {journal} {\bibinfo  {journal} {Phys.
  Rev. A}\ }\textbf {\bibinfo {volume} {96}},\ \bibinfo {pages} {022507}
  (\bibinfo {year} {2017})}\BibitemShut {NoStop}%
\bibitem [{\citenamefont {Jim{\'{e}}nez-Gal{\'{a}}n}\ \emph
  {et~al.}(2016)\citenamefont {Jim{\'{e}}nez-Gal{\'{a}}n}, \citenamefont
  {Mart{\'{i}}n},\ and\ \citenamefont {Argenti}}]{Jimenez-Galan2016}%
  \BibitemOpen
  \bibfield  {author} {\bibinfo {author} {\bibfnamefont {{\'{A}}.}~\bibnamefont
  {Jim{\'{e}}nez-Gal{\'{a}}n}}, \bibinfo {author} {\bibfnamefont
  {F.}~\bibnamefont {Mart{\'{i}}n}}, \ and\ \bibinfo {author} {\bibfnamefont
  {L.}~\bibnamefont {Argenti}},\ }\href {\doibase 10.1103/PhysRevA.93.023429}
  {\bibfield  {journal} {\bibinfo  {journal} {Phys. Rev. A}\ }\textbf {\bibinfo
  {volume} {93}},\ \bibinfo {pages} {023429} (\bibinfo {year}
  {2016})}\BibitemShut {NoStop}%
\bibitem [{\citenamefont {Gruson}\ \emph {et~al.}(2016)\citenamefont {Gruson},
  \citenamefont {Barreau}, \citenamefont {Jim{\'{e}}nez-Galan}, \citenamefont
  {Risoud}, \citenamefont {Caillat}, \citenamefont {Maquet}, \citenamefont
  {Carr{\'{e}}}, \citenamefont {Lepetit}, \citenamefont {Hergott},
  \citenamefont {Ruchon}, \citenamefont {Argenti}, \citenamefont {Ta{\"{i}}eb},
  \citenamefont {Mart{\'{i}}n},\ and\ \citenamefont
  {Sali{\`{e}}res}}]{Gruson2016}%
  \BibitemOpen
  \bibfield  {author} {\bibinfo {author} {\bibfnamefont {V.}~\bibnamefont
  {Gruson}}, \bibinfo {author} {\bibfnamefont {L.}~\bibnamefont {Barreau}},
  \bibinfo {author} {\bibnamefont {Jim{\'{e}}nez-Galan}}, \bibinfo {author}
  {\bibfnamefont {F.}~\bibnamefont {Risoud}}, \bibinfo {author} {\bibfnamefont
  {J.}~\bibnamefont {Caillat}}, \bibinfo {author} {\bibfnamefont
  {A.}~\bibnamefont {Maquet}}, \bibinfo {author} {\bibfnamefont
  {B.}~\bibnamefont {Carr{\'{e}}}}, \bibinfo {author} {\bibfnamefont
  {F.}~\bibnamefont {Lepetit}}, \bibinfo {author} {\bibfnamefont {J.~F.}\
  \bibnamefont {Hergott}}, \bibinfo {author} {\bibfnamefont {T.}~\bibnamefont
  {Ruchon}}, \bibinfo {author} {\bibfnamefont {L.}~\bibnamefont {Argenti}},
  \bibinfo {author} {\bibfnamefont {R.}~\bibnamefont {Ta{\"{i}}eb}}, \bibinfo
  {author} {\bibfnamefont {F.}~\bibnamefont {Mart{\'{i}}n}}, \ and\ \bibinfo
  {author} {\bibfnamefont {P.}~\bibnamefont {Sali{\`{e}}res}},\ }\href
  {\doibase 10.1126/science.aah5188} {\bibfield  {journal} {\bibinfo  {journal}
  {Science}\ }\textbf {\bibinfo {volume} {354}},\ \bibinfo {pages} {734}
  (\bibinfo {year} {2016})}\BibitemShut {NoStop}%
\bibitem [{\citenamefont {Kotur}\ \emph {et~al.}(2016)\citenamefont {Kotur},
  \citenamefont {Gu{\'{e}}not}, \citenamefont {Jim{\'{e}}nez-Gal{\'{a}}n},
  \citenamefont {Kroon}, \citenamefont {Larsen}, \citenamefont {Louisy},
  \citenamefont {Bengtsson}, \citenamefont {Miranda}, \citenamefont
  {Mauritsson}, \citenamefont {Arnold}, \citenamefont {Canton}, \citenamefont
  {Gisselbrecht}, \citenamefont {Carette}, \citenamefont {Dahlstr{\"{o}}m},
  \citenamefont {Lindroth}, \citenamefont {Maquet}, \citenamefont {Argenti},
  \citenamefont {Mart{\'{i}}n},\ and\ \citenamefont {L'Huillier}}]{Kotur2016}%
  \BibitemOpen
  \bibfield  {author} {\bibinfo {author} {\bibfnamefont {M.}~\bibnamefont
  {Kotur}}, \bibinfo {author} {\bibfnamefont {D.}~\bibnamefont {Gu{\'{e}}not}},
  \bibinfo {author} {\bibnamefont {Jim{\'{e}}nez-Gal{\'{a}}n}}, \bibinfo
  {author} {\bibfnamefont {D.}~\bibnamefont {Kroon}}, \bibinfo {author}
  {\bibfnamefont {E.~W.}\ \bibnamefont {Larsen}}, \bibinfo {author}
  {\bibfnamefont {M.}~\bibnamefont {Louisy}}, \bibinfo {author} {\bibfnamefont
  {S.}~\bibnamefont {Bengtsson}}, \bibinfo {author} {\bibfnamefont
  {M.}~\bibnamefont {Miranda}}, \bibinfo {author} {\bibfnamefont
  {J.}~\bibnamefont {Mauritsson}}, \bibinfo {author} {\bibfnamefont {C.~L.}\
  \bibnamefont {Arnold}}, \bibinfo {author} {\bibfnamefont {S.~E.}\
  \bibnamefont {Canton}}, \bibinfo {author} {\bibfnamefont {M.}~\bibnamefont
  {Gisselbrecht}}, \bibinfo {author} {\bibfnamefont {T.}~\bibnamefont
  {Carette}}, \bibinfo {author} {\bibfnamefont {J.~M.}\ \bibnamefont
  {Dahlstr{\"{o}}m}}, \bibinfo {author} {\bibfnamefont {E.}~\bibnamefont
  {Lindroth}}, \bibinfo {author} {\bibfnamefont {A.}~\bibnamefont {Maquet}},
  \bibinfo {author} {\bibfnamefont {L.}~\bibnamefont {Argenti}}, \bibinfo
  {author} {\bibfnamefont {F.}~\bibnamefont {Mart{\'{i}}n}}, \ and\ \bibinfo
  {author} {\bibfnamefont {A.}~\bibnamefont {L'Huillier}},\ }\href {\doibase
  10.1038/ncomms10566} {\bibfield  {journal} {\bibinfo  {journal} {Nat.
  Commun.}\ }\textbf {\bibinfo {volume} {7}},\ \bibinfo {pages} {10566}
  (\bibinfo {year} {2016})},\ \Eprint {http://arxiv.org/abs/1505.02024}
  {arXiv:1505.02024} \BibitemShut {NoStop}%
\bibitem [{\citenamefont {Stoo{\ss}}\ \emph {et~al.}(2018)\citenamefont
  {Stoo{\ss}}, \citenamefont {Cavaletto}, \citenamefont {Donsa}, \citenamefont
  {Bl{\"{a}}ttermann}, \citenamefont {Birk}, \citenamefont {Keitel},
  \citenamefont {Březinov{\'{a}}}, \citenamefont {Burgd{\"{o}}rfer},
  \citenamefont {Ott},\ and\ \citenamefont {Pfeifer}}]{Stoos2018}%
  \BibitemOpen
  \bibfield  {author} {\bibinfo {author} {\bibfnamefont {V.}~\bibnamefont
  {Stoo{\ss}}}, \bibinfo {author} {\bibfnamefont {S.~M.}\ \bibnamefont
  {Cavaletto}}, \bibinfo {author} {\bibfnamefont {S.}~\bibnamefont {Donsa}},
  \bibinfo {author} {\bibfnamefont {A.}~\bibnamefont {Bl{\"{a}}ttermann}},
  \bibinfo {author} {\bibfnamefont {P.}~\bibnamefont {Birk}}, \bibinfo {author}
  {\bibfnamefont {C.~H.}\ \bibnamefont {Keitel}}, \bibinfo {author}
  {\bibfnamefont {I.}~\bibnamefont {Březinov{\'{a}}}}, \bibinfo {author}
  {\bibfnamefont {J.}~\bibnamefont {Burgd{\"{o}}rfer}}, \bibinfo {author}
  {\bibfnamefont {C.}~\bibnamefont {Ott}}, \ and\ \bibinfo {author}
  {\bibfnamefont {T.}~\bibnamefont {Pfeifer}},\ }\href {\doibase
  10.1103/PhysRevLett.121.173005} {\bibfield  {journal} {\bibinfo  {journal}
  {Phys. Rev. Lett.}\ }\textbf {\bibinfo {volume} {121}},\ \bibinfo {pages}
  {173005} (\bibinfo {year} {2018})}\BibitemShut {NoStop}%
\bibitem [{\citenamefont {Polli}\ \emph {et~al.}(2010)\citenamefont {Polli},
  \citenamefont {Alto{\`{e}}}, \citenamefont {Weingart}, \citenamefont
  {Spillane}, \citenamefont {Manzoni}, \citenamefont {Brida}, \citenamefont
  {Tomasello}, \citenamefont {Orlandi}, \citenamefont {Kukura}, \citenamefont
  {Mathies}, \citenamefont {Garavelli},\ and\ \citenamefont
  {Cerullo}}]{Polli2010}%
  \BibitemOpen
  \bibfield  {author} {\bibinfo {author} {\bibfnamefont {D.}~\bibnamefont
  {Polli}}, \bibinfo {author} {\bibfnamefont {P.}~\bibnamefont {Alto{\`{e}}}},
  \bibinfo {author} {\bibfnamefont {O.}~\bibnamefont {Weingart}}, \bibinfo
  {author} {\bibfnamefont {K.~M.}\ \bibnamefont {Spillane}}, \bibinfo {author}
  {\bibfnamefont {C.}~\bibnamefont {Manzoni}}, \bibinfo {author} {\bibfnamefont
  {D.}~\bibnamefont {Brida}}, \bibinfo {author} {\bibfnamefont
  {G.}~\bibnamefont {Tomasello}}, \bibinfo {author} {\bibfnamefont
  {G.}~\bibnamefont {Orlandi}}, \bibinfo {author} {\bibfnamefont
  {P.}~\bibnamefont {Kukura}}, \bibinfo {author} {\bibfnamefont {R.~A.}\
  \bibnamefont {Mathies}}, \bibinfo {author} {\bibfnamefont {M.}~\bibnamefont
  {Garavelli}}, \ and\ \bibinfo {author} {\bibfnamefont {G.}~\bibnamefont
  {Cerullo}},\ }\href {\doibase 10.1038/nature09346} {\bibfield  {journal}
  {\bibinfo  {journal} {Nature}\ }\textbf {\bibinfo {volume} {467}},\ \bibinfo
  {pages} {440} (\bibinfo {year} {2010})}\BibitemShut {NoStop}%
\bibitem [{\citenamefont {Itatani}\ \emph {et~al.}(2002)\citenamefont
  {Itatani}, \citenamefont {Qu{\'{e}}r{\'{e}}}, \citenamefont {Yudin},
  \citenamefont {Ivanov}, \citenamefont {Krausz},\ and\ \citenamefont
  {Corkum}}]{Itatani2002}%
  \BibitemOpen
  \bibfield  {author} {\bibinfo {author} {\bibfnamefont {J.}~\bibnamefont
  {Itatani}}, \bibinfo {author} {\bibfnamefont {F.}~\bibnamefont
  {Qu{\'{e}}r{\'{e}}}}, \bibinfo {author} {\bibfnamefont {G.~L.}\ \bibnamefont
  {Yudin}}, \bibinfo {author} {\bibfnamefont {M.~Y.}\ \bibnamefont {Ivanov}},
  \bibinfo {author} {\bibfnamefont {F.}~\bibnamefont {Krausz}}, \ and\ \bibinfo
  {author} {\bibfnamefont {P.~B.}\ \bibnamefont {Corkum}},\ }\href {\doibase
  10.1103/PhysRevLett.88.173903} {\bibfield  {journal} {\bibinfo  {journal}
  {Phys. Rev. Lett.}\ }\textbf {\bibinfo {volume} {88}},\ \bibinfo {pages} {4}
  (\bibinfo {year} {2002})}\BibitemShut {NoStop}%
\bibitem [{\citenamefont {Ossiander}\ \emph {et~al.}(2017)\citenamefont
  {Ossiander}, \citenamefont {Siegrist}, \citenamefont {Shirvanyan},
  \citenamefont {Pazourek}, \citenamefont {Sommer}, \citenamefont {Latka},
  \citenamefont {Guggenmos}, \citenamefont {Nagele}, \citenamefont {Feist},
  \citenamefont {Burgd{\"{o}}rfer}, \citenamefont {Kienberger},\ and\
  \citenamefont {Schultze}}]{Ossiander2016}%
  \BibitemOpen
  \bibfield  {author} {\bibinfo {author} {\bibfnamefont {M.}~\bibnamefont
  {Ossiander}}, \bibinfo {author} {\bibfnamefont {F.}~\bibnamefont {Siegrist}},
  \bibinfo {author} {\bibfnamefont {V.}~\bibnamefont {Shirvanyan}}, \bibinfo
  {author} {\bibfnamefont {R.}~\bibnamefont {Pazourek}}, \bibinfo {author}
  {\bibfnamefont {A.}~\bibnamefont {Sommer}}, \bibinfo {author} {\bibfnamefont
  {T.}~\bibnamefont {Latka}}, \bibinfo {author} {\bibfnamefont
  {A.}~\bibnamefont {Guggenmos}}, \bibinfo {author} {\bibfnamefont
  {S.}~\bibnamefont {Nagele}}, \bibinfo {author} {\bibfnamefont
  {J.}~\bibnamefont {Feist}}, \bibinfo {author} {\bibfnamefont
  {J.}~\bibnamefont {Burgd{\"{o}}rfer}}, \bibinfo {author} {\bibfnamefont
  {R.}~\bibnamefont {Kienberger}}, \ and\ \bibinfo {author} {\bibfnamefont
  {M.}~\bibnamefont {Schultze}},\ }\href {\doibase 10.1038/nphys3941}
  {\bibfield  {journal} {\bibinfo  {journal} {Nat. Phys.}\ }\textbf {\bibinfo
  {volume} {13}},\ \bibinfo {pages} {280} (\bibinfo {year} {2017})}\BibitemShut
  {NoStop}%
\bibitem [{\citenamefont {Paul}\ \emph {et~al.}(2001)\citenamefont {Paul},
  \citenamefont {Toma}, \citenamefont {Breger}, \citenamefont {Mullot},
  \citenamefont {Auge}, \citenamefont {Balcou}, \citenamefont {Muller},\ and\
  \citenamefont {Agostini}}]{Paul2001}%
  \BibitemOpen
  \bibfield  {author} {\bibinfo {author} {\bibfnamefont {P.~M.}\ \bibnamefont
  {Paul}}, \bibinfo {author} {\bibfnamefont {E.~S.}\ \bibnamefont {Toma}},
  \bibinfo {author} {\bibfnamefont {P.}~\bibnamefont {Breger}}, \bibinfo
  {author} {\bibfnamefont {G.}~\bibnamefont {Mullot}}, \bibinfo {author}
  {\bibfnamefont {F.}~\bibnamefont {Auge}}, \bibinfo {author} {\bibfnamefont
  {P.}~\bibnamefont {Balcou}}, \bibinfo {author} {\bibfnamefont {H.~G.}\
  \bibnamefont {Muller}}, \ and\ \bibinfo {author} {\bibfnamefont
  {P.}~\bibnamefont {Agostini}},\ }\href {\doibase 10.1126/science.1059413}
  {\bibfield  {journal} {\bibinfo  {journal} {Science}\ }\textbf {\bibinfo
  {volume} {292}},\ \bibinfo {pages} {1689} (\bibinfo {year}
  {2001})}\BibitemShut {NoStop}%
\bibitem [{\citenamefont {Muller}(2002)}]{Muller2002}%
  \BibitemOpen
  \bibfield  {author} {\bibinfo {author} {\bibfnamefont {H.}~\bibnamefont
  {Muller}},\ }\href {\doibase 10.1007/s00340-002-0894-8} {\bibfield  {journal}
  {\bibinfo  {journal} {Appl. Phys. B}\ }\textbf {\bibinfo {volume} {74}},\
  \bibinfo {pages} {s17} (\bibinfo {year} {2002})}\BibitemShut {NoStop}%
\bibitem [{\citenamefont {Busto}\ \emph {et~al.}(2018)\citenamefont {Busto},
  \citenamefont {Barreau}, \citenamefont {Isinger}, \citenamefont {Turconi},
  \citenamefont {Alexandridi}, \citenamefont {Harth}, \citenamefont {Zhong},
  \citenamefont {Squibb}, \citenamefont {Kroon}, \citenamefont {Plogmaker},
  \citenamefont {Miranda}, \citenamefont {Jim{\'{e}}nez-Gal{\'{a}}n},
  \citenamefont {Argenti}, \citenamefont {Arnold}, \citenamefont {Feifel},
  \citenamefont {Mart{\'{i}}n}, \citenamefont {Gisselbrecht}, \citenamefont
  {L'Huillier},\ and\ \citenamefont {Sali{\`{e}}res}}]{Busto2018}%
  \BibitemOpen
  \bibfield  {author} {\bibinfo {author} {\bibfnamefont {D.}~\bibnamefont
  {Busto}}, \bibinfo {author} {\bibfnamefont {L.}~\bibnamefont {Barreau}},
  \bibinfo {author} {\bibfnamefont {M.}~\bibnamefont {Isinger}}, \bibinfo
  {author} {\bibfnamefont {M.}~\bibnamefont {Turconi}}, \bibinfo {author}
  {\bibfnamefont {C.}~\bibnamefont {Alexandridi}}, \bibinfo {author}
  {\bibfnamefont {A.}~\bibnamefont {Harth}}, \bibinfo {author} {\bibfnamefont
  {S.}~\bibnamefont {Zhong}}, \bibinfo {author} {\bibfnamefont {R.~J.}\
  \bibnamefont {Squibb}}, \bibinfo {author} {\bibfnamefont {D.}~\bibnamefont
  {Kroon}}, \bibinfo {author} {\bibfnamefont {S.}~\bibnamefont {Plogmaker}},
  \bibinfo {author} {\bibfnamefont {M.}~\bibnamefont {Miranda}}, \bibinfo
  {author} {\bibfnamefont {{\'{A}}.}~\bibnamefont {Jim{\'{e}}nez-Gal{\'{a}}n}},
  \bibinfo {author} {\bibfnamefont {L.}~\bibnamefont {Argenti}}, \bibinfo
  {author} {\bibfnamefont {C.~L.}\ \bibnamefont {Arnold}}, \bibinfo {author}
  {\bibfnamefont {R.}~\bibnamefont {Feifel}}, \bibinfo {author} {\bibfnamefont
  {F.}~\bibnamefont {Mart{\'{i}}n}}, \bibinfo {author} {\bibfnamefont
  {M.}~\bibnamefont {Gisselbrecht}}, \bibinfo {author} {\bibfnamefont
  {A.}~\bibnamefont {L'Huillier}}, \ and\ \bibinfo {author} {\bibfnamefont
  {P.}~\bibnamefont {Sali{\`{e}}res}},\ }\href {\doibase
  10.1088/1361-6455/aaa057} {\bibfield  {journal} {\bibinfo  {journal} {J.
  Phys. B At. Mol. Opt. Phys.}\ }\textbf {\bibinfo {volume} {51}},\ \bibinfo
  {pages} {044002} (\bibinfo {year} {2018})},\ \Eprint
  {http://arxiv.org/abs/1709.07639} {arXiv:1709.07639} \BibitemShut {NoStop}%
\bibitem [{\citenamefont {Barreau}\ \emph {et~al.}(2019)\citenamefont
  {Barreau}, \citenamefont {Petersson}, \citenamefont {Klinker}, \citenamefont
  {Camper}, \citenamefont {Marante}, \citenamefont {Gorman}, \citenamefont
  {Kiesewetter}, \citenamefont {Argenti}, \citenamefont {Agostini},
  \citenamefont {Gonz{\'{a}}lez-V{\'{a}}zquez}, \citenamefont {Sali{\`{e}}res},
  \citenamefont {DiMauro},\ and\ \citenamefont {Mart{\'{i}}n}}]{Barreau2019}%
  \BibitemOpen
  \bibfield  {author} {\bibinfo {author} {\bibfnamefont {L.}~\bibnamefont
  {Barreau}}, \bibinfo {author} {\bibfnamefont {C.~L.~M.}\ \bibnamefont
  {Petersson}}, \bibinfo {author} {\bibfnamefont {M.}~\bibnamefont {Klinker}},
  \bibinfo {author} {\bibfnamefont {A.}~\bibnamefont {Camper}}, \bibinfo
  {author} {\bibfnamefont {C.}~\bibnamefont {Marante}}, \bibinfo {author}
  {\bibfnamefont {T.}~\bibnamefont {Gorman}}, \bibinfo {author} {\bibfnamefont
  {D.}~\bibnamefont {Kiesewetter}}, \bibinfo {author} {\bibfnamefont
  {L.}~\bibnamefont {Argenti}}, \bibinfo {author} {\bibfnamefont
  {P.}~\bibnamefont {Agostini}}, \bibinfo {author} {\bibfnamefont
  {J.}~\bibnamefont {Gonz{\'{a}}lez-V{\'{a}}zquez}}, \bibinfo {author}
  {\bibfnamefont {P.}~\bibnamefont {Sali{\`{e}}res}}, \bibinfo {author}
  {\bibfnamefont {L.~F.}\ \bibnamefont {DiMauro}}, \ and\ \bibinfo {author}
  {\bibfnamefont {F.}~\bibnamefont {Mart{\'{i}}n}},\ }\href {\doibase
  10.1103/PhysRevLett.122.253203} {\bibfield  {journal} {\bibinfo  {journal}
  {Phys. Rev. Lett.}\ }\textbf {\bibinfo {volume} {122}},\ \bibinfo {pages}
  {253203} (\bibinfo {year} {2019})}\BibitemShut {NoStop}%
\bibitem [{\citenamefont {Donsa}\ \emph
  {et~al.}(2019{\natexlab{a}})\citenamefont {Donsa}, \citenamefont {Douguet},
  \citenamefont {Burgd{\"{o}}rfer}, \citenamefont {Březinov{\'{a}}},\ and\
  \citenamefont {Argenti}}]{Donsa2019a}%
  \BibitemOpen
  \bibfield  {author} {\bibinfo {author} {\bibfnamefont {S.}~\bibnamefont
  {Donsa}}, \bibinfo {author} {\bibfnamefont {N.}~\bibnamefont {Douguet}},
  \bibinfo {author} {\bibfnamefont {J.}~\bibnamefont {Burgd{\"{o}}rfer}},
  \bibinfo {author} {\bibfnamefont {I.}~\bibnamefont {Březinov{\'{a}}}}, \
  and\ \bibinfo {author} {\bibfnamefont {L.}~\bibnamefont {Argenti}},\ }\href
  {\doibase 10.1103/PhysRevLett.123.133203} {\bibfield  {journal} {\bibinfo
  {journal} {Phys. Rev. Lett.}\ }\textbf {\bibinfo {volume} {123}} (\bibinfo
  {year} {2019}{\natexlab{a}}),\ 10.1103/PhysRevLett.123.133203},\ \Eprint
  {http://arxiv.org/abs/1904.04380} {arXiv:1904.04380} \BibitemShut {NoStop}%
\bibitem [{\citenamefont {Laurent}\ \emph {et~al.}(2012)\citenamefont
  {Laurent}, \citenamefont {Cao}, \citenamefont {Li}, \citenamefont {Wang},
  \citenamefont {Ben-Itzhak},\ and\ \citenamefont {Cocke}}]{Laurent2012}%
  \BibitemOpen
  \bibfield  {author} {\bibinfo {author} {\bibfnamefont {G.}~\bibnamefont
  {Laurent}}, \bibinfo {author} {\bibfnamefont {W.}~\bibnamefont {Cao}},
  \bibinfo {author} {\bibfnamefont {H.}~\bibnamefont {Li}}, \bibinfo {author}
  {\bibfnamefont {Z.}~\bibnamefont {Wang}}, \bibinfo {author} {\bibfnamefont
  {I.}~\bibnamefont {Ben-Itzhak}}, \ and\ \bibinfo {author} {\bibfnamefont
  {C.~L.}\ \bibnamefont {Cocke}},\ }\href {\doibase
  10.1103/PhysRevLett.109.083001} {\bibfield  {journal} {\bibinfo  {journal}
  {Phys. Rev. Lett.}\ }\textbf {\bibinfo {volume} {109}},\ \bibinfo {pages}
  {083001} (\bibinfo {year} {2012})}\BibitemShut {NoStop}%
\bibitem [{\citenamefont {Laurent}\ \emph {et~al.}(2013)\citenamefont
  {Laurent}, \citenamefont {Cao}, \citenamefont {Ben-Itzhak},\ and\
  \citenamefont {Cocke}}]{Laurent2013}%
  \BibitemOpen
  \bibfield  {author} {\bibinfo {author} {\bibfnamefont {G.}~\bibnamefont
  {Laurent}}, \bibinfo {author} {\bibfnamefont {W.}~\bibnamefont {Cao}},
  \bibinfo {author} {\bibfnamefont {I.}~\bibnamefont {Ben-Itzhak}}, \ and\
  \bibinfo {author} {\bibfnamefont {C.~L.}\ \bibnamefont {Cocke}},\ }\href
  {\doibase 10.1364/OE.21.016914} {\bibfield  {journal} {\bibinfo  {journal}
  {Opt. Express}\ }\textbf {\bibinfo {volume} {21}},\ \bibinfo {pages} {16914}
  (\bibinfo {year} {2013})}\BibitemShut {NoStop}%
\bibitem [{\citenamefont {Loriot}\ \emph {et~al.}(2017)\citenamefont {Loriot},
  \citenamefont {Marciniak}, \citenamefont {Karras}, \citenamefont {Schindler},
  \citenamefont {Renois-Predelus}, \citenamefont {Compagnon}, \citenamefont
  {Concina}, \citenamefont {Br{\'{e}}dy}, \citenamefont {Celep}, \citenamefont
  {Bordas}, \citenamefont {Constant},\ and\ \citenamefont
  {L{\'{e}}pine}}]{Loriot2017}%
  \BibitemOpen
  \bibfield  {author} {\bibinfo {author} {\bibfnamefont {V.}~\bibnamefont
  {Loriot}}, \bibinfo {author} {\bibfnamefont {A.}~\bibnamefont {Marciniak}},
  \bibinfo {author} {\bibfnamefont {G.}~\bibnamefont {Karras}}, \bibinfo
  {author} {\bibfnamefont {B.}~\bibnamefont {Schindler}}, \bibinfo {author}
  {\bibfnamefont {G.}~\bibnamefont {Renois-Predelus}}, \bibinfo {author}
  {\bibfnamefont {I.}~\bibnamefont {Compagnon}}, \bibinfo {author}
  {\bibfnamefont {B.}~\bibnamefont {Concina}}, \bibinfo {author} {\bibfnamefont
  {R.}~\bibnamefont {Br{\'{e}}dy}}, \bibinfo {author} {\bibfnamefont
  {G.}~\bibnamefont {Celep}}, \bibinfo {author} {\bibfnamefont
  {C.}~\bibnamefont {Bordas}}, \bibinfo {author} {\bibfnamefont
  {E.}~\bibnamefont {Constant}}, \ and\ \bibinfo {author} {\bibfnamefont
  {F.}~\bibnamefont {L{\'{e}}pine}},\ }\href {\doibase
  10.1088/2040-8986/aa8e10} {\bibfield  {journal} {\bibinfo  {journal} {J. Opt.
  (United Kingdom)}\ }\textbf {\bibinfo {volume} {19}},\ \bibinfo {pages}
  {114003} (\bibinfo {year} {2017})}\BibitemShut {NoStop}%
\bibitem [{\citenamefont {Loriot}\ \emph {et~al.}(2020)\citenamefont {Loriot},
  \citenamefont {Marciniak}, \citenamefont {Nandi}, \citenamefont {Karras},
  \citenamefont {Herv{\'{e}}}, \citenamefont {Constant}, \citenamefont
  {Pl{\'{e}}siat}, \citenamefont {Palacios}, \citenamefont {Mart{\'{i}}n},\
  and\ \citenamefont {L{\'{e}}pine}}]{Loriot2020}%
  \BibitemOpen
  \bibfield  {author} {\bibinfo {author} {\bibfnamefont {V.}~\bibnamefont
  {Loriot}}, \bibinfo {author} {\bibfnamefont {A.}~\bibnamefont {Marciniak}},
  \bibinfo {author} {\bibfnamefont {S.}~\bibnamefont {Nandi}}, \bibinfo
  {author} {\bibfnamefont {G.}~\bibnamefont {Karras}}, \bibinfo {author}
  {\bibfnamefont {M.}~\bibnamefont {Herv{\'{e}}}}, \bibinfo {author}
  {\bibfnamefont {E.}~\bibnamefont {Constant}}, \bibinfo {author}
  {\bibfnamefont {E.}~\bibnamefont {Pl{\'{e}}siat}}, \bibinfo {author}
  {\bibfnamefont {A.}~\bibnamefont {Palacios}}, \bibinfo {author}
  {\bibfnamefont {F.}~\bibnamefont {Mart{\'{i}}n}}, \ and\ \bibinfo {author}
  {\bibfnamefont {F.}~\bibnamefont {L{\'{e}}pine}},\ }\href {\doibase
  10.1088/2515-7647/ab7b10} {\bibfield  {journal} {\bibinfo  {journal} {J.
  Phys. Photonics}\ }\textbf {\bibinfo {volume} {2}},\ \bibinfo {pages}
  {024003} (\bibinfo {year} {2020})}\BibitemShut {NoStop}%
\bibitem [{\citenamefont {Shin}\ \emph {et~al.}(1999)\citenamefont {Shin},
  \citenamefont {Lee}, \citenamefont {Cha}, \citenamefont {Hong},\ and\
  \citenamefont {Nam}}]{Shin1999}%
  \BibitemOpen
  \bibfield  {author} {\bibinfo {author} {\bibfnamefont {H.~J.}\ \bibnamefont
  {Shin}}, \bibinfo {author} {\bibfnamefont {D.~G.}\ \bibnamefont {Lee}},
  \bibinfo {author} {\bibfnamefont {Y.~H.}\ \bibnamefont {Cha}}, \bibinfo
  {author} {\bibfnamefont {K.~H.}\ \bibnamefont {Hong}}, \ and\ \bibinfo
  {author} {\bibfnamefont {C.~H.}\ \bibnamefont {Nam}},\ }\href {\doibase
  10.1103/PhysRevLett.83.2544} {\bibfield  {journal} {\bibinfo  {journal}
  {Phys. Rev. Lett.}\ }\textbf {\bibinfo {volume} {83}},\ \bibinfo {pages}
  {2544} (\bibinfo {year} {1999})}\BibitemShut {NoStop}%
\bibitem [{\citenamefont {Shin}\ \emph {et~al.}(2001)\citenamefont {Shin},
  \citenamefont {Lee}, \citenamefont {Cha}, \citenamefont {Kim}, \citenamefont
  {Hong},\ and\ \citenamefont {Nam}}]{Shin2001}%
  \BibitemOpen
  \bibfield  {author} {\bibinfo {author} {\bibfnamefont {H.~J.}\ \bibnamefont
  {Shin}}, \bibinfo {author} {\bibfnamefont {D.~G.}\ \bibnamefont {Lee}},
  \bibinfo {author} {\bibfnamefont {Y.~H.}\ \bibnamefont {Cha}}, \bibinfo
  {author} {\bibfnamefont {J.~H.}\ \bibnamefont {Kim}}, \bibinfo {author}
  {\bibfnamefont {K.~H.}\ \bibnamefont {Hong}}, \ and\ \bibinfo {author}
  {\bibfnamefont {C.~H.}\ \bibnamefont {Nam}},\ }\href {\doibase
  10.1103/PhysRevA.63.053407} {\bibfield  {journal} {\bibinfo  {journal} {Phys.
  Rev. A - At. Mol. Opt. Phys.}\ }\textbf {\bibinfo {volume} {63}},\ \bibinfo
  {pages} {9} (\bibinfo {year} {2001})}\BibitemShut {NoStop}%
\bibitem [{\citenamefont {Mairesse}\ \emph {et~al.}(2005)\citenamefont
  {Mairesse}, \citenamefont {Gobert}, \citenamefont {Breger}, \citenamefont
  {Merdji}, \citenamefont {Meynadier}, \citenamefont {Monchicourt},
  \citenamefont {Perdrix}, \citenamefont {Sali{\`{e}}res},\ and\ \citenamefont
  {Carr{\'{e}}}}]{Mairesse2005}%
  \BibitemOpen
  \bibfield  {author} {\bibinfo {author} {\bibfnamefont {Y.}~\bibnamefont
  {Mairesse}}, \bibinfo {author} {\bibfnamefont {O.}~\bibnamefont {Gobert}},
  \bibinfo {author} {\bibfnamefont {P.}~\bibnamefont {Breger}}, \bibinfo
  {author} {\bibfnamefont {H.}~\bibnamefont {Merdji}}, \bibinfo {author}
  {\bibfnamefont {P.}~\bibnamefont {Meynadier}}, \bibinfo {author}
  {\bibfnamefont {P.}~\bibnamefont {Monchicourt}}, \bibinfo {author}
  {\bibfnamefont {M.}~\bibnamefont {Perdrix}}, \bibinfo {author} {\bibfnamefont
  {P.}~\bibnamefont {Sali{\`{e}}res}}, \ and\ \bibinfo {author} {\bibfnamefont
  {B.}~\bibnamefont {Carr{\'{e}}}},\ }\href {\doibase
  10.1103/PhysRevLett.94.173903} {\bibfield  {journal} {\bibinfo  {journal}
  {Phys. Rev. Lett.}\ }\textbf {\bibinfo {volume} {94}},\ \bibinfo {pages}
  {173903} (\bibinfo {year} {2005})}\BibitemShut {NoStop}%
\bibitem [{\citenamefont {Haessler}\ \emph {et~al.}(2012)\citenamefont
  {Haessler}, \citenamefont {Bom}, \citenamefont {Gobert}, \citenamefont
  {Hergott}, \citenamefont {Lepetit}, \citenamefont {Perdrix}, \citenamefont
  {Carr{\'{e}}}, \citenamefont {Ozaki},\ and\ \citenamefont
  {Sali{\`{e}}res}}]{Haessler2012}%
  \BibitemOpen
  \bibfield  {author} {\bibinfo {author} {\bibfnamefont {S.}~\bibnamefont
  {Haessler}}, \bibinfo {author} {\bibfnamefont {L.~B.}\ \bibnamefont {Bom}},
  \bibinfo {author} {\bibfnamefont {O.}~\bibnamefont {Gobert}}, \bibinfo
  {author} {\bibfnamefont {J.~F.}\ \bibnamefont {Hergott}}, \bibinfo {author}
  {\bibfnamefont {F.}~\bibnamefont {Lepetit}}, \bibinfo {author} {\bibfnamefont
  {M.}~\bibnamefont {Perdrix}}, \bibinfo {author} {\bibfnamefont
  {B.}~\bibnamefont {Carr{\'{e}}}}, \bibinfo {author} {\bibfnamefont
  {T.}~\bibnamefont {Ozaki}}, \ and\ \bibinfo {author} {\bibfnamefont
  {P.}~\bibnamefont {Sali{\`{e}}res}},\ }\href {\doibase
  10.1088/0953-4075/45/7/074012} {\bibfield  {journal} {\bibinfo  {journal} {J.
  Phys. B At. Mol. Opt. Phys.}\ }\textbf {\bibinfo {volume} {45}},\ \bibinfo
  {pages} {074012} (\bibinfo {year} {2012})}\BibitemShut {NoStop}%
\bibitem [{\citenamefont {Ardana-Lamas}\ \emph {et~al.}(2016)\citenamefont
  {Ardana-Lamas}, \citenamefont {Erny}, \citenamefont {Stepanov}, \citenamefont
  {Gorgisyan}, \citenamefont {Jurani{\'{c}}}, \citenamefont {Abela},\ and\
  \citenamefont {Hauri}}]{Ardana-Lamas2016}%
  \BibitemOpen
  \bibfield  {author} {\bibinfo {author} {\bibfnamefont {F.}~\bibnamefont
  {Ardana-Lamas}}, \bibinfo {author} {\bibfnamefont {C.}~\bibnamefont {Erny}},
  \bibinfo {author} {\bibfnamefont {A.~G.}\ \bibnamefont {Stepanov}}, \bibinfo
  {author} {\bibfnamefont {I.}~\bibnamefont {Gorgisyan}}, \bibinfo {author}
  {\bibfnamefont {P.}~\bibnamefont {Jurani{\'{c}}}}, \bibinfo {author}
  {\bibfnamefont {R.}~\bibnamefont {Abela}}, \ and\ \bibinfo {author}
  {\bibfnamefont {C.~P.}\ \bibnamefont {Hauri}},\ }\href {\doibase
  10.1103/PhysRevA.93.043838} {\bibfield  {journal} {\bibinfo  {journal} {Phys.
  Rev. A}\ }\textbf {\bibinfo {volume} {93}},\ \bibinfo {pages} {043838}
  (\bibinfo {year} {2016})},\ \Eprint {http://arxiv.org/abs/1504.01958}
  {arXiv:1504.01958} \BibitemShut {NoStop}%
\bibitem [{\citenamefont {Kim}\ \emph {et~al.}(2013)\citenamefont {Kim},
  \citenamefont {Zhang}, \citenamefont {Ruchon}, \citenamefont {Hergott},
  \citenamefont {Auguste}, \citenamefont {Villeneuve}, \citenamefont {Corkum},\
  and\ \citenamefont {Qu{\'{e}}r{\'{e}}}}]{Kim2013}%
  \BibitemOpen
  \bibfield  {author} {\bibinfo {author} {\bibfnamefont {K.~T.}\ \bibnamefont
  {Kim}}, \bibinfo {author} {\bibfnamefont {C.}~\bibnamefont {Zhang}}, \bibinfo
  {author} {\bibfnamefont {T.}~\bibnamefont {Ruchon}}, \bibinfo {author}
  {\bibfnamefont {J.~F.}\ \bibnamefont {Hergott}}, \bibinfo {author}
  {\bibfnamefont {T.}~\bibnamefont {Auguste}}, \bibinfo {author} {\bibfnamefont
  {D.~M.}\ \bibnamefont {Villeneuve}}, \bibinfo {author} {\bibfnamefont
  {P.~B.}\ \bibnamefont {Corkum}}, \ and\ \bibinfo {author} {\bibfnamefont
  {F.}~\bibnamefont {Qu{\'{e}}r{\'{e}}}},\ }\href {\doibase
  10.1038/nphoton.2013.170} {\bibfield  {journal} {\bibinfo  {journal} {Nat.
  Photonics}\ }\textbf {\bibinfo {volume} {7}},\ \bibinfo {pages} {651}
  (\bibinfo {year} {2013})}\BibitemShut {NoStop}%
\bibitem [{\citenamefont {Varj{\'{u}}}\ \emph {et~al.}(2005)\citenamefont
  {Varj{\'{u}}}, \citenamefont {Mairesse}, \citenamefont {Carr{\'{e}}},
  \citenamefont {Gaarde}, \citenamefont {Johnsson}, \citenamefont {Kazamias},
  \citenamefont {L{\'{o}}pez-Martens}, \citenamefont {Mauritsson},
  \citenamefont {Schafer}, \citenamefont {Balcou}, \citenamefont {L'Huillier},\
  and\ \citenamefont {Sali{\`{e}}res}}]{Varju2005}%
  \BibitemOpen
  \bibfield  {author} {\bibinfo {author} {\bibfnamefont {K.}~\bibnamefont
  {Varj{\'{u}}}}, \bibinfo {author} {\bibfnamefont {Y.}~\bibnamefont
  {Mairesse}}, \bibinfo {author} {\bibfnamefont {B.}~\bibnamefont
  {Carr{\'{e}}}}, \bibinfo {author} {\bibfnamefont {M.~B.}\ \bibnamefont
  {Gaarde}}, \bibinfo {author} {\bibfnamefont {P.}~\bibnamefont {Johnsson}},
  \bibinfo {author} {\bibfnamefont {S.}~\bibnamefont {Kazamias}}, \bibinfo
  {author} {\bibfnamefont {R.}~\bibnamefont {L{\'{o}}pez-Martens}}, \bibinfo
  {author} {\bibfnamefont {J.}~\bibnamefont {Mauritsson}}, \bibinfo {author}
  {\bibfnamefont {K.~J.}\ \bibnamefont {Schafer}}, \bibinfo {author}
  {\bibfnamefont {P.}~\bibnamefont {Balcou}}, \bibinfo {author} {\bibfnamefont
  {A.}~\bibnamefont {L'Huillier}}, \ and\ \bibinfo {author} {\bibfnamefont
  {P.}~\bibnamefont {Sali{\`{e}}res}},\ }in\ \href {\doibase
  10.1080/09500340412331301542} {\emph {\bibinfo {booktitle} {J. Mod. Opt.}}},\
  Vol.~\bibinfo {volume} {52}\ (\bibinfo  {publisher} {Taylor {\&} Francis
  Group},\ \bibinfo {year} {2005})\ pp.\ \bibinfo {pages}
  {379--394}\BibitemShut {NoStop}%
\bibitem [{\citenamefont {Mauritsson}\ \emph {et~al.}(2004)\citenamefont
  {Mauritsson}, \citenamefont {Johnsson}, \citenamefont {L{\'{o}}pez-Martens},
  \citenamefont {Varj{\'{u}}}, \citenamefont {Kornelis}, \citenamefont
  {Biegert}, \citenamefont {Keller}, \citenamefont {Gaarde}, \citenamefont
  {Schafer},\ and\ \citenamefont {L'Huillier}}]{Mauritsson2004}%
  \BibitemOpen
  \bibfield  {author} {\bibinfo {author} {\bibfnamefont {J.}~\bibnamefont
  {Mauritsson}}, \bibinfo {author} {\bibfnamefont {P.}~\bibnamefont
  {Johnsson}}, \bibinfo {author} {\bibfnamefont {R.}~\bibnamefont
  {L{\'{o}}pez-Martens}}, \bibinfo {author} {\bibfnamefont {K.}~\bibnamefont
  {Varj{\'{u}}}}, \bibinfo {author} {\bibfnamefont {W.}~\bibnamefont
  {Kornelis}}, \bibinfo {author} {\bibfnamefont {J.}~\bibnamefont {Biegert}},
  \bibinfo {author} {\bibfnamefont {U.}~\bibnamefont {Keller}}, \bibinfo
  {author} {\bibfnamefont {M.~B.}\ \bibnamefont {Gaarde}}, \bibinfo {author}
  {\bibfnamefont {K.~J.}\ \bibnamefont {Schafer}}, \ and\ \bibinfo {author}
  {\bibfnamefont {A.}~\bibnamefont {L'Huillier}},\ }\href {\doibase
  10.1103/PhysRevA.70.021801} {\bibfield  {journal} {\bibinfo  {journal} {Phys.
  Rev. A}\ }\textbf {\bibinfo {volume} {70}},\ \bibinfo {pages} {021801}
  (\bibinfo {year} {2004})}\BibitemShut {NoStop}%
\bibitem [{\citenamefont {Wigner}(1955)}]{Wigner1955}%
  \BibitemOpen
  \bibfield  {author} {\bibinfo {author} {\bibfnamefont {E.~P.}\ \bibnamefont
  {Wigner}},\ }\href {\doibase 10.1103/PhysRev.98.145} {\bibfield  {journal}
  {\bibinfo  {journal} {Phys. Rev.}\ }\textbf {\bibinfo {volume} {98}},\
  \bibinfo {pages} {145} (\bibinfo {year} {1955})}\BibitemShut {NoStop}%
\bibitem [{\citenamefont {Smith}(1960)}]{Smith1960}%
  \BibitemOpen
  \bibfield  {author} {\bibinfo {author} {\bibfnamefont {F.~T.}\ \bibnamefont
  {Smith}},\ }\href {\doibase 10.1103/PhysRev.118.349} {\bibfield  {journal}
  {\bibinfo  {journal} {Phys. Rev.}\ }\textbf {\bibinfo {volume} {118}},\
  \bibinfo {pages} {349} (\bibinfo {year} {1960})}\BibitemShut {NoStop}%
\bibitem [{\citenamefont {Nagele}\ \emph {et~al.}(2011)\citenamefont {Nagele},
  \citenamefont {Pazourek}, \citenamefont {Feist}, \citenamefont
  {Doblhoff-Dier}, \citenamefont {Lemell}, \citenamefont {Tők{\'{e}}si},\ and\
  \citenamefont {Burgd{\"{o}}rfer}}]{Nagele2011a}%
  \BibitemOpen
  \bibfield  {author} {\bibinfo {author} {\bibfnamefont {S.}~\bibnamefont
  {Nagele}}, \bibinfo {author} {\bibfnamefont {R.}~\bibnamefont {Pazourek}},
  \bibinfo {author} {\bibfnamefont {J.}~\bibnamefont {Feist}}, \bibinfo
  {author} {\bibfnamefont {K.}~\bibnamefont {Doblhoff-Dier}}, \bibinfo {author}
  {\bibfnamefont {C.}~\bibnamefont {Lemell}}, \bibinfo {author} {\bibfnamefont
  {K.}~\bibnamefont {Tők{\'{e}}si}}, \ and\ \bibinfo {author} {\bibfnamefont
  {J.}~\bibnamefont {Burgd{\"{o}}rfer}},\ }\href {\doibase
  10.1088/0953-4075/44/8/081001} {\bibfield  {journal} {\bibinfo  {journal} {J.
  Phys. B At. Mol. Opt. Phys.}\ }\textbf {\bibinfo {volume} {44}},\ \bibinfo
  {pages} {081001} (\bibinfo {year} {2011})}\BibitemShut {NoStop}%
\bibitem [{\citenamefont {Kl{\"{u}}nder}\ \emph {et~al.}(2011)\citenamefont
  {Kl{\"{u}}nder}, \citenamefont {Dahlstr{\"{o}}m}, \citenamefont
  {Gisselbrecht}, \citenamefont {Fordell}, \citenamefont {Swoboda},
  \citenamefont {Gu{\'{e}}not}, \citenamefont {Johnsson}, \citenamefont
  {Caillat}, \citenamefont {Mauritsson}, \citenamefont {Maquet}, \citenamefont
  {Ta{\"{i}}eb},\ and\ \citenamefont {L'Huillier}}]{Klunder2011}%
  \BibitemOpen
  \bibfield  {author} {\bibinfo {author} {\bibfnamefont {K.}~\bibnamefont
  {Kl{\"{u}}nder}}, \bibinfo {author} {\bibfnamefont {J.~M.}\ \bibnamefont
  {Dahlstr{\"{o}}m}}, \bibinfo {author} {\bibfnamefont {M.}~\bibnamefont
  {Gisselbrecht}}, \bibinfo {author} {\bibfnamefont {T.}~\bibnamefont
  {Fordell}}, \bibinfo {author} {\bibfnamefont {M.}~\bibnamefont {Swoboda}},
  \bibinfo {author} {\bibfnamefont {D.}~\bibnamefont {Gu{\'{e}}not}}, \bibinfo
  {author} {\bibfnamefont {P.}~\bibnamefont {Johnsson}}, \bibinfo {author}
  {\bibfnamefont {J.}~\bibnamefont {Caillat}}, \bibinfo {author} {\bibfnamefont
  {J.}~\bibnamefont {Mauritsson}}, \bibinfo {author} {\bibfnamefont
  {A.}~\bibnamefont {Maquet}}, \bibinfo {author} {\bibfnamefont
  {R.}~\bibnamefont {Ta{\"{i}}eb}}, \ and\ \bibinfo {author} {\bibfnamefont
  {A.}~\bibnamefont {L'Huillier}},\ }\href {\doibase
  10.1103/PhysRevLett.106.143002} {\bibfield  {journal} {\bibinfo  {journal}
  {Phys. Rev. Lett.}\ }\textbf {\bibinfo {volume} {106}},\ \bibinfo {pages}
  {143002} (\bibinfo {year} {2011})}\BibitemShut {NoStop}%
\bibitem [{\citenamefont {Mairesse}\ \emph {et~al.}(2003)\citenamefont
  {Mairesse}, \citenamefont {de~Bohan}, \citenamefont {Frasinski},
  \citenamefont {Merdji}, \citenamefont {Dinu}, \citenamefont {Monchicourt},
  \citenamefont {Breger}, \citenamefont {Kovacev}, \citenamefont {Taieb},
  \citenamefont {Carre}, \citenamefont {Muller}, \citenamefont {Agostini},\
  and\ \citenamefont {Salieres}}]{Mairesse2003}%
  \BibitemOpen
  \bibfield  {author} {\bibinfo {author} {\bibfnamefont {Y.}~\bibnamefont
  {Mairesse}}, \bibinfo {author} {\bibfnamefont {D.}~\bibnamefont {de~Bohan}},
  \bibinfo {author} {\bibfnamefont {L.~J.}\ \bibnamefont {Frasinski}}, \bibinfo
  {author} {\bibfnamefont {H.}~\bibnamefont {Merdji}}, \bibinfo {author}
  {\bibfnamefont {L.~C.}\ \bibnamefont {Dinu}}, \bibinfo {author}
  {\bibfnamefont {P.}~\bibnamefont {Monchicourt}}, \bibinfo {author}
  {\bibfnamefont {P.}~\bibnamefont {Breger}}, \bibinfo {author} {\bibfnamefont
  {M.}~\bibnamefont {Kovacev}}, \bibinfo {author} {\bibfnamefont
  {R.}~\bibnamefont {Taieb}}, \bibinfo {author} {\bibfnamefont
  {B.}~\bibnamefont {Carre}}, \bibinfo {author} {\bibfnamefont {H.~G.}\
  \bibnamefont {Muller}}, \bibinfo {author} {\bibfnamefont {P.}~\bibnamefont
  {Agostini}}, \ and\ \bibinfo {author} {\bibfnamefont {P.}~\bibnamefont
  {Salieres}},\ }\href {\doibase 10.1126/science.1090277} {\bibfield  {journal}
  {\bibinfo  {journal} {Science}\ }\textbf {\bibinfo {volume} {302}},\ \bibinfo
  {pages} {1540} (\bibinfo {year} {2003})}\BibitemShut {NoStop}%
\bibitem [{\citenamefont {V{\'{e}}niard}\ \emph {et~al.}(1996)\citenamefont
  {V{\'{e}}niard}, \citenamefont {Ta{\"{i}}eb},\ and\ \citenamefont
  {Maquet}}]{Veniard1996}%
  \BibitemOpen
  \bibfield  {author} {\bibinfo {author} {\bibfnamefont {V.}~\bibnamefont
  {V{\'{e}}niard}}, \bibinfo {author} {\bibfnamefont {R.}~\bibnamefont
  {Ta{\"{i}}eb}}, \ and\ \bibinfo {author} {\bibfnamefont {A.}~\bibnamefont
  {Maquet}},\ }\href {\doibase 10.1103/PhysRevA.54.721} {\bibfield  {journal}
  {\bibinfo  {journal} {Phys. Rev. A}\ }\textbf {\bibinfo {volume} {54}},\
  \bibinfo {pages} {721} (\bibinfo {year} {1996})}\BibitemShut {NoStop}%
\bibitem [{\citenamefont {Dahlstr{\"{o}}m}\ \emph {et~al.}(2013)\citenamefont
  {Dahlstr{\"{o}}m}, \citenamefont {Gu{\'{e}}not}, \citenamefont
  {Kl{\"{u}}nder}, \citenamefont {Gisselbrecht}, \citenamefont {Mauritsson},
  \citenamefont {L'Huillier}, \citenamefont {Maquet},\ and\ \citenamefont
  {Ta{\"{i}}eb}}]{Dahlstrom2013}%
  \BibitemOpen
  \bibfield  {author} {\bibinfo {author} {\bibfnamefont {J.~M.}\ \bibnamefont
  {Dahlstr{\"{o}}m}}, \bibinfo {author} {\bibfnamefont {D.}~\bibnamefont
  {Gu{\'{e}}not}}, \bibinfo {author} {\bibfnamefont {K.}~\bibnamefont
  {Kl{\"{u}}nder}}, \bibinfo {author} {\bibfnamefont {M.}~\bibnamefont
  {Gisselbrecht}}, \bibinfo {author} {\bibfnamefont {J.}~\bibnamefont
  {Mauritsson}}, \bibinfo {author} {\bibfnamefont {A.}~\bibnamefont
  {L'Huillier}}, \bibinfo {author} {\bibfnamefont {A.}~\bibnamefont {Maquet}},
  \ and\ \bibinfo {author} {\bibfnamefont {R.}~\bibnamefont {Ta{\"{i}}eb}},\
  }\href {\doibase 10.1016/j.chemphys.2012.01.017} {\bibfield  {journal}
  {\bibinfo  {journal} {Chem. Phys.}\ }\textbf {\bibinfo {volume} {414}},\
  \bibinfo {pages} {53} (\bibinfo {year} {2013})}\BibitemShut {NoStop}%
\bibitem [{\citenamefont {Fuchs}\ \emph {et~al.}(2020)\citenamefont {Fuchs},
  \citenamefont {Douguet}, \citenamefont {Donsa}, \citenamefont {Martin},
  \citenamefont {Burgd{\"{o}}rfer}, \citenamefont {Argenti}, \citenamefont
  {Cattaneo},\ and\ \citenamefont {Keller}}]{Fuchs2020}%
  \BibitemOpen
  \bibfield  {author} {\bibinfo {author} {\bibfnamefont {J.}~\bibnamefont
  {Fuchs}}, \bibinfo {author} {\bibfnamefont {N.}~\bibnamefont {Douguet}},
  \bibinfo {author} {\bibfnamefont {S.}~\bibnamefont {Donsa}}, \bibinfo
  {author} {\bibfnamefont {F.}~\bibnamefont {Martin}}, \bibinfo {author}
  {\bibfnamefont {J.}~\bibnamefont {Burgd{\"{o}}rfer}}, \bibinfo {author}
  {\bibfnamefont {L.}~\bibnamefont {Argenti}}, \bibinfo {author} {\bibfnamefont
  {L.}~\bibnamefont {Cattaneo}}, \ and\ \bibinfo {author} {\bibfnamefont
  {U.}~\bibnamefont {Keller}},\ }\href {\doibase 10.1364/optica.378639}
  {\bibfield  {journal} {\bibinfo  {journal} {Optica}\ }\textbf {\bibinfo
  {volume} {7}},\ \bibinfo {pages} {154} (\bibinfo {year} {2020})},\ \Eprint
  {http://arxiv.org/abs/1907.03607} {arXiv:1907.03607} \BibitemShut {NoStop}%
\bibitem [{\citenamefont {Busto}\ \emph {et~al.}(2019)\citenamefont {Busto},
  \citenamefont {Vinbladh}, \citenamefont {Zhong}, \citenamefont {Isinger},
  \citenamefont {Nandi}, \citenamefont {Maclot}, \citenamefont {Johnsson},
  \citenamefont {Gisselbrecht}, \citenamefont {L'Huillier}, \citenamefont
  {Lindroth},\ and\ \citenamefont {Dahlstr{\"{o}}m}}]{Busto2019}%
  \BibitemOpen
  \bibfield  {author} {\bibinfo {author} {\bibfnamefont {D.}~\bibnamefont
  {Busto}}, \bibinfo {author} {\bibfnamefont {J.}~\bibnamefont {Vinbladh}},
  \bibinfo {author} {\bibfnamefont {S.}~\bibnamefont {Zhong}}, \bibinfo
  {author} {\bibfnamefont {M.}~\bibnamefont {Isinger}}, \bibinfo {author}
  {\bibfnamefont {S.}~\bibnamefont {Nandi}}, \bibinfo {author} {\bibfnamefont
  {S.}~\bibnamefont {Maclot}}, \bibinfo {author} {\bibfnamefont
  {P.}~\bibnamefont {Johnsson}}, \bibinfo {author} {\bibfnamefont
  {M.}~\bibnamefont {Gisselbrecht}}, \bibinfo {author} {\bibfnamefont
  {A.}~\bibnamefont {L'Huillier}}, \bibinfo {author} {\bibfnamefont
  {E.}~\bibnamefont {Lindroth}}, \ and\ \bibinfo {author} {\bibfnamefont
  {J.~M.}\ \bibnamefont {Dahlstr{\"{o}}m}},\ }\href {\doibase
  10.1103/PhysRevLett.123.133201} {\bibfield  {journal} {\bibinfo  {journal}
  {Phys. Rev. Lett.}\ }\textbf {\bibinfo {volume} {123}},\ \bibinfo {pages}
  {133201} (\bibinfo {year} {2019})},\ \Eprint
  {http://arxiv.org/abs/1811.05341} {arXiv:1811.05341} \BibitemShut {NoStop}%
\bibitem [{\citenamefont {Tong}\ and\ \citenamefont {Lin}(2005)}]{Tong2005}%
  \BibitemOpen
  \bibfield  {author} {\bibinfo {author} {\bibfnamefont {X.~M.}\ \bibnamefont
  {Tong}}\ and\ \bibinfo {author} {\bibfnamefont {C.~D.}\ \bibnamefont {Lin}},\
  }\href {\doibase 10.1088/0953-4075/38/15/001} {\bibfield  {journal} {\bibinfo
   {journal} {J. Phys. B At. Mol. Opt. Phys.}\ }\textbf {\bibinfo {volume}
  {38}},\ \bibinfo {pages} {2593} (\bibinfo {year} {2005})}\BibitemShut
  {NoStop}%
\bibitem [{\citenamefont {D{\"{o}}rner}\ \emph {et~al.}(2000)\citenamefont
  {D{\"{o}}rner}, \citenamefont {Mergel}, \citenamefont {Jagutzki},
  \citenamefont {Spielberger}, \citenamefont {Ullrich}, \citenamefont
  {Moshammer},\ and\ \citenamefont {Schmidt-B{\"{o}}cking}}]{Dorner2000}%
  \BibitemOpen
  \bibfield  {author} {\bibinfo {author} {\bibfnamefont {R.}~\bibnamefont
  {D{\"{o}}rner}}, \bibinfo {author} {\bibfnamefont {V.}~\bibnamefont
  {Mergel}}, \bibinfo {author} {\bibfnamefont {O.}~\bibnamefont {Jagutzki}},
  \bibinfo {author} {\bibfnamefont {L.}~\bibnamefont {Spielberger}}, \bibinfo
  {author} {\bibfnamefont {J.}~\bibnamefont {Ullrich}}, \bibinfo {author}
  {\bibfnamefont {R.}~\bibnamefont {Moshammer}}, \ and\ \bibinfo {author}
  {\bibfnamefont {H.}~\bibnamefont {Schmidt-B{\"{o}}cking}},\ }\href {\doibase
  10.1016/S0370-1573(99)00109-X} {\bibfield  {journal} {\bibinfo  {journal}
  {Phys. Rep.}\ }\textbf {\bibinfo {volume} {330}},\ \bibinfo {pages} {95}
  (\bibinfo {year} {2000})}\BibitemShut {NoStop}%
\bibitem [{\citenamefont {Heuser}\ \emph {et~al.}(2016)\citenamefont {Heuser},
  \citenamefont {{Jim{\'{e}}nez Gal{\'{a}}n}}, \citenamefont {Cirelli},
  \citenamefont {Marante}, \citenamefont {Sabbar}, \citenamefont {Boge},
  \citenamefont {Lucchini}, \citenamefont {Gallmann}, \citenamefont {Ivanov},
  \citenamefont {Kheifets}, \citenamefont {Dahlstr{\"{o}}m}, \citenamefont
  {Lindroth}, \citenamefont {Argenti}, \citenamefont {Mart{\'{i}}n},\ and\
  \citenamefont {Keller}}]{Heuser2016}%
  \BibitemOpen
  \bibfield  {author} {\bibinfo {author} {\bibfnamefont {S.}~\bibnamefont
  {Heuser}}, \bibinfo {author} {\bibfnamefont {{\'{A}}.}~\bibnamefont
  {{Jim{\'{e}}nez Gal{\'{a}}n}}}, \bibinfo {author} {\bibfnamefont
  {C.}~\bibnamefont {Cirelli}}, \bibinfo {author} {\bibfnamefont
  {C.}~\bibnamefont {Marante}}, \bibinfo {author} {\bibfnamefont
  {M.}~\bibnamefont {Sabbar}}, \bibinfo {author} {\bibfnamefont
  {R.}~\bibnamefont {Boge}}, \bibinfo {author} {\bibfnamefont {M.}~\bibnamefont
  {Lucchini}}, \bibinfo {author} {\bibfnamefont {L.}~\bibnamefont {Gallmann}},
  \bibinfo {author} {\bibfnamefont {I.}~\bibnamefont {Ivanov}}, \bibinfo
  {author} {\bibfnamefont {A.~S.}\ \bibnamefont {Kheifets}}, \bibinfo {author}
  {\bibfnamefont {J.~M.}\ \bibnamefont {Dahlstr{\"{o}}m}}, \bibinfo {author}
  {\bibfnamefont {E.}~\bibnamefont {Lindroth}}, \bibinfo {author}
  {\bibfnamefont {L.}~\bibnamefont {Argenti}}, \bibinfo {author} {\bibfnamefont
  {F.}~\bibnamefont {Mart{\'{i}}n}}, \ and\ \bibinfo {author} {\bibfnamefont
  {U.}~\bibnamefont {Keller}},\ }\href {\doibase 10.1103/PhysRevA.94.063409}
  {\bibfield  {journal} {\bibinfo  {journal} {Phys. Rev. A}\ }\textbf {\bibinfo
  {volume} {94}},\ \bibinfo {pages} {063409} (\bibinfo {year}
  {2016})}\BibitemShut {NoStop}%
\bibitem [{\citenamefont {Sabbar}\ \emph {et~al.}(2014)\citenamefont {Sabbar},
  \citenamefont {Heuser}, \citenamefont {Boge}, \citenamefont {Lucchini},
  \citenamefont {Gallmann}, \citenamefont {Cirelli},\ and\ \citenamefont
  {Keller}}]{Sabbar2014}%
  \BibitemOpen
  \bibfield  {author} {\bibinfo {author} {\bibfnamefont {M.}~\bibnamefont
  {Sabbar}}, \bibinfo {author} {\bibfnamefont {S.}~\bibnamefont {Heuser}},
  \bibinfo {author} {\bibfnamefont {R.}~\bibnamefont {Boge}}, \bibinfo {author}
  {\bibfnamefont {M.}~\bibnamefont {Lucchini}}, \bibinfo {author}
  {\bibfnamefont {L.}~\bibnamefont {Gallmann}}, \bibinfo {author}
  {\bibfnamefont {C.}~\bibnamefont {Cirelli}}, \ and\ \bibinfo {author}
  {\bibfnamefont {U.}~\bibnamefont {Keller}},\ }\href {\doibase
  10.1063/1.4898017} {\bibfield  {journal} {\bibinfo  {journal} {Rev. Sci.
  Instrum.}\ }\textbf {\bibinfo {volume} {85}},\ \bibinfo {pages} {103113}
  (\bibinfo {year} {2014})}\BibitemShut {NoStop}%
\bibitem [{\citenamefont {Douguet}\ \emph {et~al.}(2016)\citenamefont
  {Douguet}, \citenamefont {Grum-Grzhimailo}, \citenamefont {Gryzlova},
  \citenamefont {Staroselskaya}, \citenamefont {Venzke},\ and\ \citenamefont
  {Bartschat}}]{Douguet2016}%
  \BibitemOpen
  \bibfield  {author} {\bibinfo {author} {\bibfnamefont {N.}~\bibnamefont
  {Douguet}}, \bibinfo {author} {\bibfnamefont {A.~N.}\ \bibnamefont
  {Grum-Grzhimailo}}, \bibinfo {author} {\bibfnamefont {E.~V.}\ \bibnamefont
  {Gryzlova}}, \bibinfo {author} {\bibfnamefont {E.~I.}\ \bibnamefont
  {Staroselskaya}}, \bibinfo {author} {\bibfnamefont {J.}~\bibnamefont
  {Venzke}}, \ and\ \bibinfo {author} {\bibfnamefont {K.}~\bibnamefont
  {Bartschat}},\ }\href {\doibase 10.1103/PhysRevA.93.033402} {\bibfield
  {journal} {\bibinfo  {journal} {Phys. Rev. A}\ }\textbf {\bibinfo {volume}
  {93}},\ \bibinfo {pages} {033402} (\bibinfo {year} {2016})}\BibitemShut
  {NoStop}%
\bibitem [{\citenamefont {Feist}\ \emph {et~al.}(2008)\citenamefont {Feist},
  \citenamefont {Nagele}, \citenamefont {Pazourek}, \citenamefont {Persson},
  \citenamefont {Schneider}, \citenamefont {Collins},\ and\ \citenamefont
  {Burgd{\"{o}}rfer}}]{Feist2008a}%
  \BibitemOpen
  \bibfield  {author} {\bibinfo {author} {\bibfnamefont {J.}~\bibnamefont
  {Feist}}, \bibinfo {author} {\bibfnamefont {S.}~\bibnamefont {Nagele}},
  \bibinfo {author} {\bibfnamefont {R.}~\bibnamefont {Pazourek}}, \bibinfo
  {author} {\bibfnamefont {E.}~\bibnamefont {Persson}}, \bibinfo {author}
  {\bibfnamefont {B.~I.}\ \bibnamefont {Schneider}}, \bibinfo {author}
  {\bibfnamefont {L.~A.}\ \bibnamefont {Collins}}, \ and\ \bibinfo {author}
  {\bibfnamefont {J.}~\bibnamefont {Burgd{\"{o}}rfer}},\ }\href {\doibase
  10.1103/PhysRevA.77.043420} {\bibfield  {journal} {\bibinfo  {journal} {Phys.
  Rev. A}\ }\textbf {\bibinfo {volume} {77}},\ \bibinfo {pages} {043420}
  (\bibinfo {year} {2008})},\ \Eprint {http://arxiv.org/abs/0803.0511}
  {arXiv:0803.0511} \BibitemShut {NoStop}%
\bibitem [{\citenamefont {Donsa}\ \emph
  {et~al.}(2019{\natexlab{b}})\citenamefont {Donsa}, \citenamefont
  {Březinov{\'{a}}}, \citenamefont {Ni}, \citenamefont {Feist},\ and\
  \citenamefont {Burgd{\"{o}}rfer}}]{Donsa2019b}%
  \BibitemOpen
  \bibfield  {author} {\bibinfo {author} {\bibfnamefont {S.}~\bibnamefont
  {Donsa}}, \bibinfo {author} {\bibfnamefont {I.}~\bibnamefont
  {Březinov{\'{a}}}}, \bibinfo {author} {\bibfnamefont {H.}~\bibnamefont
  {Ni}}, \bibinfo {author} {\bibfnamefont {J.}~\bibnamefont {Feist}}, \ and\
  \bibinfo {author} {\bibfnamefont {J.}~\bibnamefont {Burgd{\"{o}}rfer}},\
  }\href {\doibase 10.1103/PhysRevA.99.023413} {\bibfield  {journal} {\bibinfo
  {journal} {Phys. Rev. A}\ }\textbf {\bibinfo {volume} {99}},\ \bibinfo
  {pages} {023413} (\bibinfo {year} {2019}{\natexlab{b}})},\ \Eprint
  {http://arxiv.org/abs/1811.09110} {arXiv:1811.09110} \BibitemShut {NoStop}%
\bibitem [{\citenamefont {Lewenstein}\ \emph {et~al.}(1995)\citenamefont
  {Lewenstein}, \citenamefont {Sali{\`{e}}res},\ and\ \citenamefont
  {L'Huillier}}]{Lewenstein1995}%
  \BibitemOpen
  \bibfield  {author} {\bibinfo {author} {\bibfnamefont {M.}~\bibnamefont
  {Lewenstein}}, \bibinfo {author} {\bibfnamefont {P.}~\bibnamefont
  {Sali{\`{e}}res}}, \ and\ \bibinfo {author} {\bibfnamefont {A.}~\bibnamefont
  {L'Huillier}},\ }\href {\doibase 10.1103/PhysRevA.52.4747} {\bibfield
  {journal} {\bibinfo  {journal} {Phys. Rev. A}\ }\textbf {\bibinfo {volume}
  {52}},\ \bibinfo {pages} {4747} (\bibinfo {year} {1995})}\BibitemShut
  {NoStop}%
\bibitem [{\citenamefont {Gaarde}\ and\ \citenamefont
  {Schafer}(2002)}]{Gaarde2002}%
  \BibitemOpen
  \bibfield  {author} {\bibinfo {author} {\bibfnamefont {M.~B.}\ \bibnamefont
  {Gaarde}}\ and\ \bibinfo {author} {\bibfnamefont {K.~J.}\ \bibnamefont
  {Schafer}},\ }\href {\doibase 10.1103/PhysRevA.65.031406} {\bibfield
  {journal} {\bibinfo  {journal} {Phys. Rev. A - At. Mol. Opt. Phys.}\ }\textbf
  {\bibinfo {volume} {65}},\ \bibinfo {pages} {4} (\bibinfo {year}
  {2002})}\BibitemShut {NoStop}%
\bibitem [{\citenamefont {You}\ \emph {et~al.}(2020)\citenamefont {You},
  \citenamefont {Ueda}, \citenamefont {Gryzlova}, \citenamefont
  {Grum-Grzhimailo}, \citenamefont {Popova}, \citenamefont {Staroselskaya},
  \citenamefont {Tugs}, \citenamefont {Orimo}, \citenamefont {Sato},
  \citenamefont {Ishikawa}, \citenamefont {Carpeggiani}, \citenamefont
  {Csizmadia}, \citenamefont {F{\"{u}}le}, \citenamefont {Sansone},
  \citenamefont {Maroju}, \citenamefont {D'Elia}, \citenamefont {Mazza},
  \citenamefont {Meyer}, \citenamefont {Callegari}, \citenamefont {{Di Fraia}},
  \citenamefont {Plekan}, \citenamefont {Richter}, \citenamefont {Giannessi},
  \citenamefont {Allaria}, \citenamefont {{De Ninno}}, \citenamefont
  {Trov{\`{o}}}, \citenamefont {Badano}, \citenamefont {Diviacco},
  \citenamefont {Gaio}, \citenamefont {Gauthier}, \citenamefont {Mirian},
  \citenamefont {Penco}, \citenamefont {Ribi{\v{c}}}, \citenamefont
  {Spampinati}, \citenamefont {Spezzani},\ and\ \citenamefont
  {Prince}}]{You2020}%
  \BibitemOpen
  \bibfield  {author} {\bibinfo {author} {\bibfnamefont {D.}~\bibnamefont
  {You}}, \bibinfo {author} {\bibfnamefont {K.}~\bibnamefont {Ueda}}, \bibinfo
  {author} {\bibfnamefont {E.~V.}\ \bibnamefont {Gryzlova}}, \bibinfo {author}
  {\bibfnamefont {A.~N.}\ \bibnamefont {Grum-Grzhimailo}}, \bibinfo {author}
  {\bibfnamefont {M.~M.}\ \bibnamefont {Popova}}, \bibinfo {author}
  {\bibfnamefont {E.~I.}\ \bibnamefont {Staroselskaya}}, \bibinfo {author}
  {\bibfnamefont {O.}~\bibnamefont {Tugs}}, \bibinfo {author} {\bibfnamefont
  {Y.}~\bibnamefont {Orimo}}, \bibinfo {author} {\bibfnamefont
  {T.}~\bibnamefont {Sato}}, \bibinfo {author} {\bibfnamefont {K.~L.}\
  \bibnamefont {Ishikawa}}, \bibinfo {author} {\bibfnamefont {P.~A.}\
  \bibnamefont {Carpeggiani}}, \bibinfo {author} {\bibfnamefont
  {T.}~\bibnamefont {Csizmadia}}, \bibinfo {author} {\bibfnamefont
  {M.}~\bibnamefont {F{\"{u}}le}}, \bibinfo {author} {\bibfnamefont
  {G.}~\bibnamefont {Sansone}}, \bibinfo {author} {\bibfnamefont {P.~K.}\
  \bibnamefont {Maroju}}, \bibinfo {author} {\bibfnamefont {A.}~\bibnamefont
  {D'Elia}}, \bibinfo {author} {\bibfnamefont {T.}~\bibnamefont {Mazza}},
  \bibinfo {author} {\bibfnamefont {M.}~\bibnamefont {Meyer}}, \bibinfo
  {author} {\bibfnamefont {C.}~\bibnamefont {Callegari}}, \bibinfo {author}
  {\bibfnamefont {M.}~\bibnamefont {{Di Fraia}}}, \bibinfo {author}
  {\bibfnamefont {O.}~\bibnamefont {Plekan}}, \bibinfo {author} {\bibfnamefont
  {R.}~\bibnamefont {Richter}}, \bibinfo {author} {\bibfnamefont
  {L.}~\bibnamefont {Giannessi}}, \bibinfo {author} {\bibfnamefont
  {E.}~\bibnamefont {Allaria}}, \bibinfo {author} {\bibfnamefont
  {G.}~\bibnamefont {{De Ninno}}}, \bibinfo {author} {\bibfnamefont
  {M.}~\bibnamefont {Trov{\`{o}}}}, \bibinfo {author} {\bibfnamefont
  {L.}~\bibnamefont {Badano}}, \bibinfo {author} {\bibfnamefont
  {B.}~\bibnamefont {Diviacco}}, \bibinfo {author} {\bibfnamefont
  {G.}~\bibnamefont {Gaio}}, \bibinfo {author} {\bibfnamefont {D.}~\bibnamefont
  {Gauthier}}, \bibinfo {author} {\bibfnamefont {N.}~\bibnamefont {Mirian}},
  \bibinfo {author} {\bibfnamefont {G.}~\bibnamefont {Penco}}, \bibinfo
  {author} {\bibfnamefont {P.~R.}\ \bibnamefont {Ribi{\v{c}}}}, \bibinfo
  {author} {\bibfnamefont {S.}~\bibnamefont {Spampinati}}, \bibinfo {author}
  {\bibfnamefont {C.}~\bibnamefont {Spezzani}}, \ and\ \bibinfo {author}
  {\bibfnamefont {K.~C.}\ \bibnamefont {Prince}},\ }\href {\doibase
  10.1103/physrevx.10.031070} {\bibfield  {journal} {\bibinfo  {journal} {Phys.
  Rev. X}\ }\textbf {\bibinfo {volume} {10}},\ \bibinfo {pages} {31070}
  (\bibinfo {year} {2020})}\BibitemShut {NoStop}%
\end{thebibliography}%

\foreach \x in {1,2}
{%
\clearpage


\section*{Supplementary material}

# SUPPLEMENTAL MATERIAL

## A. Characterization of the cc-transition rates

To characterize the cc-transition rates, we compare the results of numerically calculated photoelectron spectra (PES) in helium for three different cases: (1) one narrow XUV harmonic in the absence of the IR. (2) The same XUV harmonic plus an IR field with intensity $1 \cdot 10^{11} \text{W/cm}^2$. (3) The same XUV harmonic plus an IR field with intensity $2 \cdot 10^{11} \text{W/cm}^2$. The calculations are performed using the singe-active electron (SAE) approximation and the model potential from [1]. We have verified that full 2-electron simulations give nearly identical results. Since the amplitude of the IR field is constant across a time interval wider than the APT duration, and the target does not exhibit any resonance in the spectral region of interest, the results are the same as for a purely monochromatic IR light. In particular, to the lowest perturbative order, there is no quantum path interference and the resulting PES does not depend on the XUV-IR delay. We integrate the resulting total yield for the XUV-only simulation and the two-photon peaks for the two-color simulations. Fig. 1 shows the two-color PES for an XUV energy of 35 eV and IR intensity of $2 \cdot 10^{11} \text{W/cm}^2$. The two-photon peaks are well separated from the one-photon peak.

The modulus of the one-photon and two-photon-amplitudes can be obtained by taking the square root of the corresponding one-photon and two-photon yields, respectively. The cc-transition rates for absorption (emission) are then calculated by dividing the higher (lower) two-photon probability by the one-photon probability. The calculation is repeated for all XUV energies from 24.5 eV to 40 eV in steps of 0.5 eV. Fig. 2 shows the transition rates as a function of the energy, justifying the assumption of linear transition rates (Eq. (6) and (7)).

As can be seen from the comparison between the transitions rates from different IR-intensities, the transition rates for both absorption and emission scale with the square root of the IR intensity, in line with the fact that the two-photon amplitude is proportional to the electric field strength.

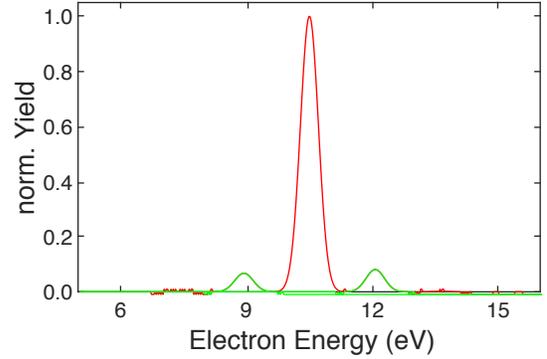

FIG. 1. (a) Photoelectron yield from the SAE simulation for helium for the two-color case with an XUV energy of 35 eV and IR intensity of $2 \cdot 10^{11} \text{W/cm}^2$. The side peaks (green) correspond to the two-photon transitions for additional absorption or emission of one IR photon.

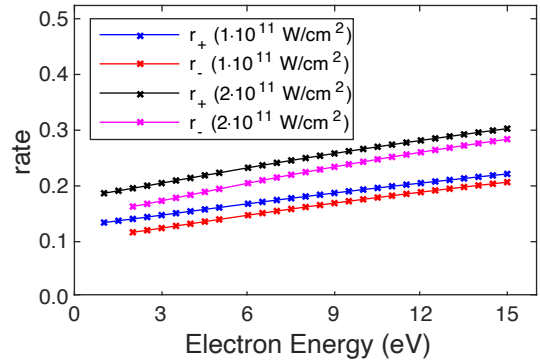

FIG. 2. Transition rates for the IR induced cc-transitions as a function of energy for additional absorption and emission of an IR photon and for different intensities of the IR field ($10^{11}$ W/cm$^2$ and $2 \cdot 10^{11}$ W/cm$^2$). The transitions rates for the higher intensity case are $\sqrt{2}$ higher compared to the weak intensity case as they scale with the strength of the electric field.

## B. Total Photoelectron yield

The total photoelectron signal is determined by integration of Eq. (2) over all emission angles and thus reads

$$\begin{aligned} f_{tot}(E,\tau) &= \int_0^{2\pi} \int_0^{\pi} I(E,\vartheta,\tau) \sin(\vartheta) d\vartheta d\phi \\ &= |A^1(E)|^2 + \sum_{\ell=s,d} [|A_\ell^+(E)|^2 + |A_\ell^-(E)|^2 \\ &\quad + 2|A_\ell^+(E)||A_\ell^-(E)| \cos(2\omega\tau + \varphi_\ell^+ - \varphi_\ell^-)] \\ &\approx |A^1(E)|^2 + 2|A^+(E)|^2 + 2|A^-(E)|^2 \\ &\quad + 4|A^+(E)||A^-(E)| \cos(2\omega\tau + \varphi^+ - \varphi^-). \end{aligned}$$

The photoelectron spectrum corresponds to the $2\omega_{\text{IR}}$-RABBITT signal. [2]. When the delay is integrated over a full IR cycle, the cosine term vanishes.

## C. Solution for the Ionization Phase

The modulus $A(E)$ and the phase $\delta(E)$ of the $\omega_{\text{IR}}$ oscillation amplitude are obtained via Fourier-transform of the experimental asymmetry signal. The quantities $a^+(E)$ and $a^-(E)$ (compare with Eq. (5) in the main text) are obtained by fitting the transition rates to the integrated PES. From the two equations

$$A(E) = \left|a^+ e^{i\Delta\varphi^+_{1-2}} - a^- e^{i\Delta\varphi^-_{1-2}}\right|,$$
$$\delta(E) = \arg\left(a^+ e^{i\Delta\varphi^+_{1-2}} - a^- e^{i\Delta\varphi^-_{1-2}}\right),$$

it is possible to retrieve $\Delta\varphi^\pm_{1-2}(E)$ analytically. Let $\chi = \Delta\varphi^-_{1-2} - \Delta\varphi^+_{1-2}$, then

$$A^2(E) = a^{+2} + a^{-2} - 2a^+ a^- \cos\chi,$$

which can be solved for $\chi$ as

$$\chi = \pm \arccos\left(\frac{A^2 - a^{+2} - a^{+2}}{2a^+ a^-}\right).$$

The correct sign determination for $\chi$ is ascertained *a posteriori* by requiring the constraint $\Delta\varphi^+_{1-2}(E) = \Delta\varphi^-_{1-2}(E - E_0)$ to be satisfied for consistency. From the value of $\chi$ and the expression for $\delta(E)$, it is straightforward to retrieve the two phases $\Delta\varphi^\pm_{1-2}$ as

$$\Delta\varphi^+_{1-2} = \delta(E) - \arg\left(a^+ - a^- e^{i\chi}\right),$$
$$\Delta\varphi^-_{1-2} = \chi + \Delta\varphi^+_{1-2}.$$

This solution holds for all energies, so that the phase differences $\Delta\varphi^\pm_{1-2}$ can be retrieved across the full spectrum.

---

[1] X. M. Tong and C. D. Lin, J. Phys. B At. Mol. Opt. Phys. **38**, 2593 (2005).
[2] J. M. Dahlström, D. Guénot, K. Klünder, M. Gisselbrecht, J. Mauritsson, A. L'Huillier, A. Maquet, and R. Taïeb, Chem. Phys. **414**, 53 (2013).


}


\end{document}